\documentclass[aps,prd, twocolumn, superscriptaddress, nofootinbib, floatfix]{revtex4-1}
\usepackage{xcolor}
\usepackage{graphicx}
\usepackage{wrapfig}
\usepackage{amsmath}
\usepackage{amsfonts}
\usepackage{amssymb}
\usepackage{multirow}
\usepackage{slashed}
\usepackage{physics}
\usepackage{float}
\usepackage[colorlinks=true, linkcolor=blue, citecolor=blue, urlcolor=blue]{hyperref}
\newcommand\befs{\begin{figure*}}
\newcommand\eefs[1]{\label{fig:#1}\end{figure*}}
\newcommand\bef{\begin{figure}}
\newcommand\eef[1]{\label{fig:#1}\end{figure}}
\newcommand\beq{\begin{equation}}
\newcommand\eeq[1]{\label{#1}\end{equation}}
\newcommand\beqa{\begin{eqnarray}}
\newcommand\eeqa[1]{\label{#1}\end{eqnarray}}
\newcommand\bet{\begin{table}}
\newcommand\eet[1]{\label{tb:#1}\end{table}}
\newcommand\bets{\begin{table*}}
\newcommand\eets[1]{\label{tb:#1}\end{table*}}
\newcommand\fgn[1]{Fig.\ \ref{fig:#1}}
\newcommand\eqn[1]{Eq.\ (\ref{#1})}

\begin{document}

\widetext

\title{Towards studying the structural differences between the pion and its radial excitation}
\author{Xiang Gao}
\email{xgao@bnl.gov}
\affiliation{Physics Department, Brookhaven National Laboratory, Upton, NY 11973, USA}
\affiliation{Physics Department, Tsinghua University, Beijing 100084, China}
\author{Nikhil Karthik}
\email{nkarthik.work@gmail.com}
\affiliation{Department of Physics, College of William \& Mary, Williamsburg, VA 23185, USA}
\affiliation{Thomas Jefferson National Accelerator Facility, Newport News, VA 23606, USA}
\author{Swagato Mukherjee}
\affiliation{Physics Department, Brookhaven National Laboratory, Upton, NY 11973, USA}
\author{Peter Petreczky}
\affiliation{Physics Department, Brookhaven National Laboratory, Upton, NY 11973, USA}
\author{Sergey Syritsyn}
\affiliation{Department of Physics and Astronomy, Stony Brook University, Stony Brook, NY 11794, USA}
\affiliation{RIKEN-BNL Research Center, Brookhaven National Lab, Upton, NY, 11973, USA}
\author{Yong Zhao}
\affiliation{Physics Department, Brookhaven National Laboratory, Upton, NY 11973, USA}

\begin{abstract}

We present an exploratory lattice QCD investigation of the differences
between the valence quark structure of pion and its radial excitation
$\pi(1300)$ in a fixed finite volume using the leading-twist
factorization approach.  We present evidences that the first pion
excitation in our lattice computation is a single particle state
that is likely to be the finite volume realization of $\pi(1300)$.
An analysis with reasonable priors result in better estimates of
the excited state PDF and the moments, wherein we find evidence
that the radial excitation of pion correlates with an almost two-fold
increase in the momentum fraction of valence quarks.  This
proof-of-principle work establishes the viability of future lattice
computations incorporating larger operator basis that can resolve
the structural changes accompanying hadronic excitation.

\end{abstract}

\date{\today}
\maketitle

\section{Introduction}

The parton structure of pion has garnered both
experimental~\cite{Badier:1983mj,Betev:1985pf,Conway:1989fs,Owens:1984zj,Sutton:1991ay,Gluck:1991ey,
Gluck:1999xe, Wijesooriya:2005ir,Barry:2018ort,Novikov:2020snp} as
well as theoretical
efforts~\cite{Aicher:2010cb,Nguyen:2011jy,Chen:2016sno,Cui:2020tdf,Roberts:2020udq,deTeramond:2018ecg,RuizArriola:2002wr,Broniowski:2017wbr,Lan:2019rba,Bednar:2018mtf}.
A better determination of the quark structure of pion is also part
of the goals of upcoming experimental
facilities~\cite{Aguilar:2019teb,Denisov:2018unj}. In addition to experimental determinations, due to the recent
breakthroughs in computing parton structure using the Euclidean
lattice QCD simulations via leading-twist perturbative factorization
approaches (Large Momentum Effective Theory~\cite{Ji:2013dva,Ji:2014gla},
short-distance factorization of the pseudo distribution~\cite{Radyushkin:2017cyf,Orginos:2017kos},
current-current correlators ~\cite{Braun:2007wv,Ma:2014jla,Ma:2017pxb}, which has also been dubbed as good lattice cross sections~\cite{Ma:2017pxb}, 
and Refs~\cite{Constantinou:2020pek,Zhao:2018fyu,Cichy:2018mum,Monahan:2018euv,Ji:2020ect}
for extensive reviews on the methodology), the valence quark structure
of pion has also been investigated from first-principle QCD
computations~\cite{Gao:2020ito,Chen:2018fwa,Izubuchi:2019lyk,Joo:2019bzr,Lin:2020ssv,Sufian:2019bol,Sufian:2020vzb,Karthik:2021qwz}.
The large-$x$ behavior of the valence pion PDF has been an unresolved
issue that has been approached using all the above lines of attack,
with the promise of being settled in the near future by lattice
computations with finer lattices, realistic physical pion masses
and with the usage of highly boosted pion states to reduce higher-twist
effects that might be amplified~\cite{Braun:2018brg, Liu:2020rqi,
Ji:2020brr} near $x=1$.

The considerable interest in the quark structure of pion is due to its
special role as the Nambu-Goldstone boson of chiral symmetry breaking
in QCD. The grand goal of this research direction is to understand
the aspects of mass-gap generation in QCD via the quark-gluon
interaction within the pion. The large-$x$ behavior of pion PDF has
been proposed to hold the key to make this connection
(c.f.,~\cite{Roberts:2020hiw}).  While the enigmatic aspect of QCD
is the presence of nonvanishing mass-gap between the vacuum and the
ground-states of various quantum numbers even in the chiral limit
(except the pseudo-scalar, which is an exception), it is equally
enigmatic that there are non-zero mass-gaps amongst the excited
states in the tower of excited spectrum as well. To contrast, if the
trace-anomaly was absent in QCD, there would
not be mass-gaps between the vacuum and the various ground-states, nor
between the excited states.
Given the stark dissimilarity between the vanishing mass of a pion in the
chiral limit and the nonvanishing masses of its excited states in
the same limit, it is reasonable to expect any differences between
the quark and gluon structures of the ground-state pion and its
excited states could help us understand the mechanism behind
spontaneous symmetry-breaking and the mass-gap generation better.
In this respect, there have been prior lattice computations to study
the decay constants of the pion and its
radial excitation~\cite{McNeile:2006qy,Mastropas:2014fsa}, where the decay constant 
of the radial excitation is expected to vanish in the chiral limit 
unlike that of the ground-state pion~\cite{Holl:2004fr}.  Closely related to the decay constant, 
the distribution amplitudes of the pion and its radial excitation have 
also been previously studied using the Dyson-Schwinger Equation~\cite{Li:2016dzv}.
With the lattice computation of PDFs now possible, a novel theoretical
research direction to study not just the differences between the
long-distance behaviors of the ground and the excited states, but
to study the differences in their internal structural properties
is promising. In this respect, we should also point to a 
previous study~\cite{Chai:2020nxw} of the $\Delta^+$ 
baryon on the lattice, which differs by both mass and angular momentum from that of 
the proton. Since there is also experimental thrust to understand
exotic gluon excitations of mesons in Jefferson Lab 12 GeV
program~\cite{Dudek:2012vr}, studies as the present one on the
parton structure of simpler radial excitations, might be
helpful phenomenologically by providing a case to contrast the exotic transitions with.
\befs
\centering
\includegraphics[scale=0.46]{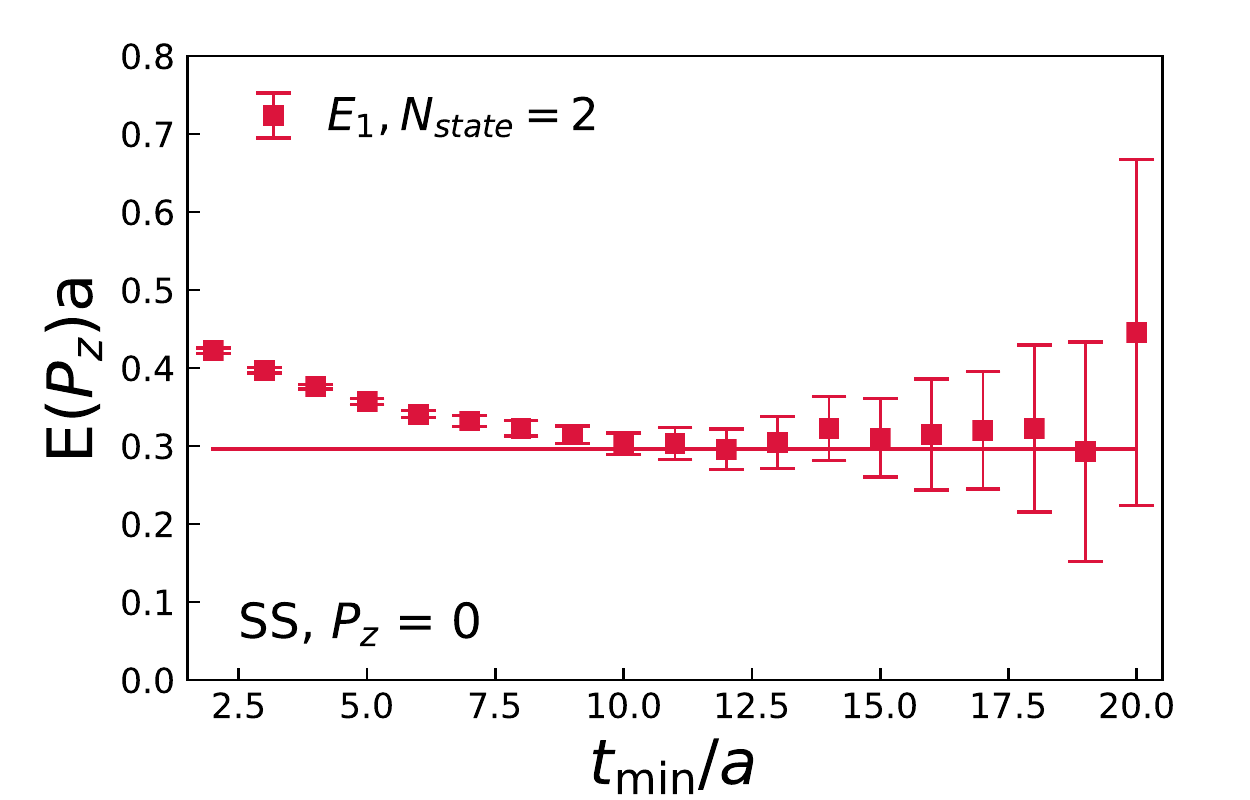}
\includegraphics[scale=0.46]{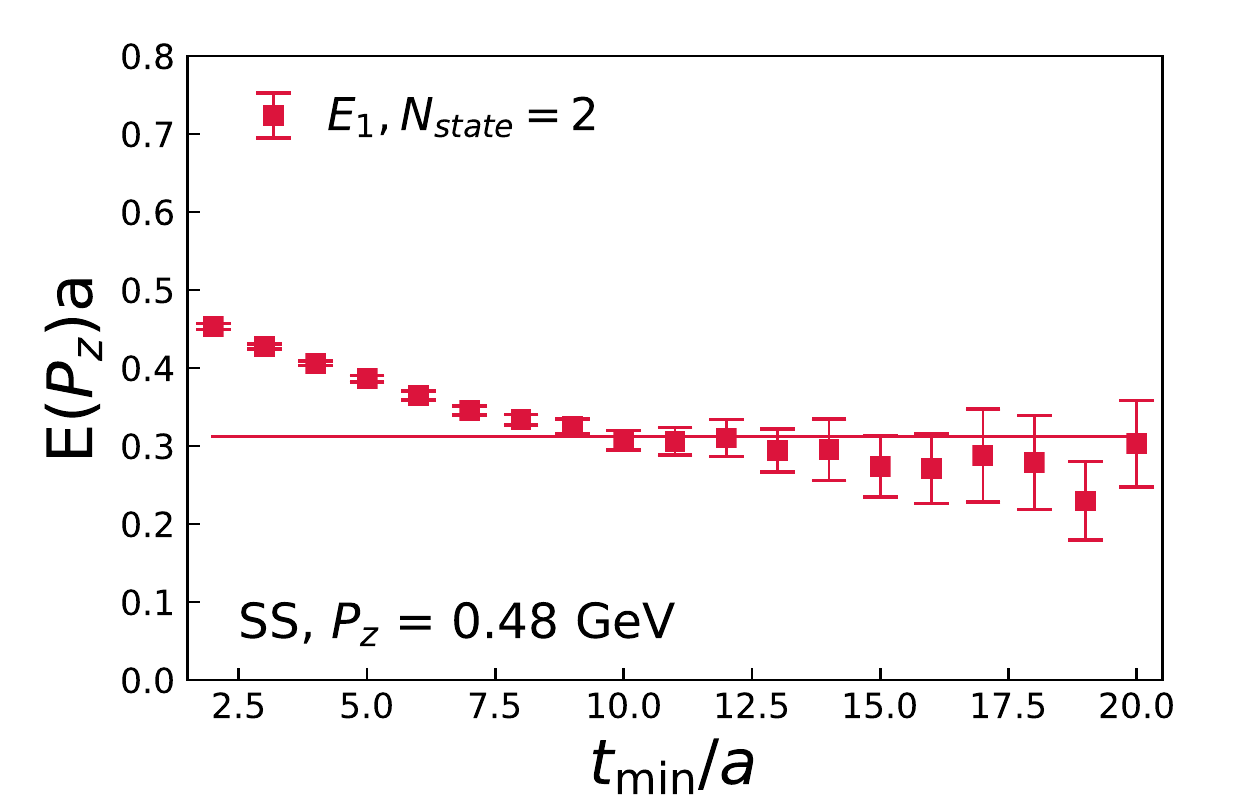}
\includegraphics[scale=0.46]{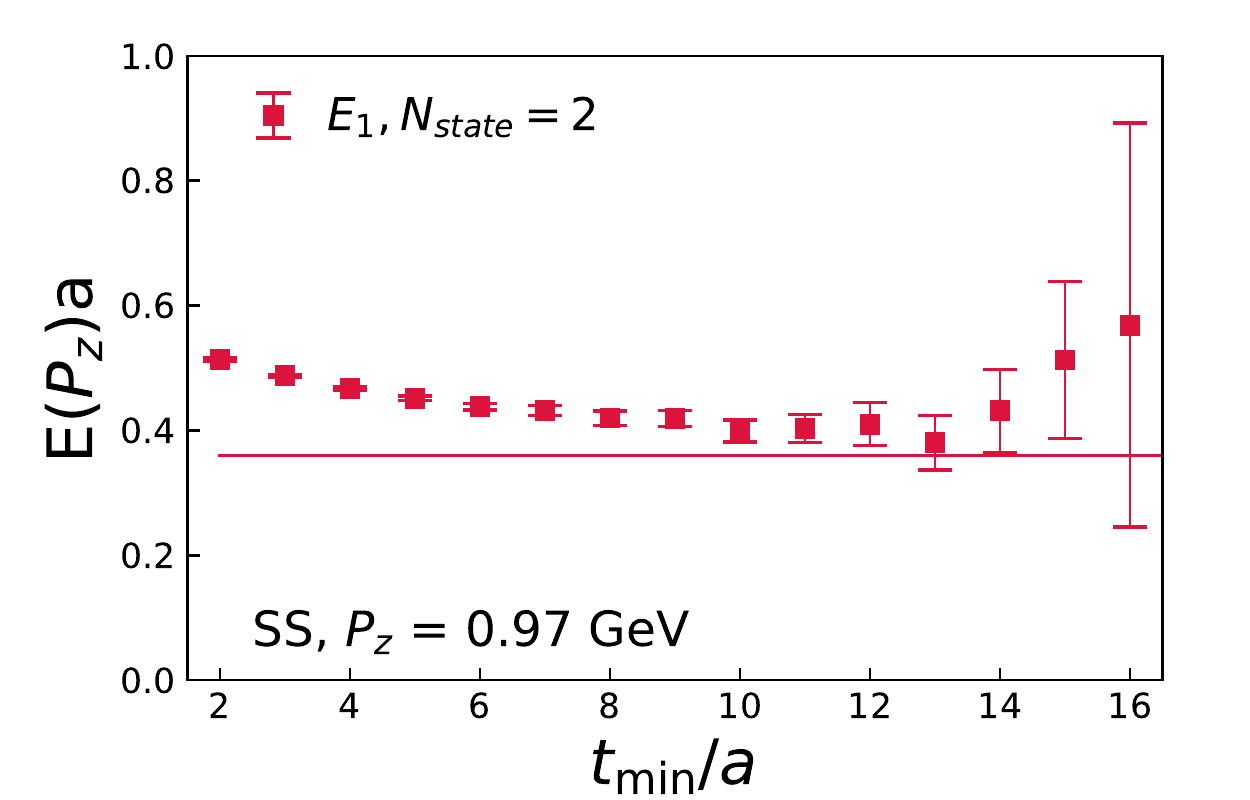}
\includegraphics[scale=0.46]{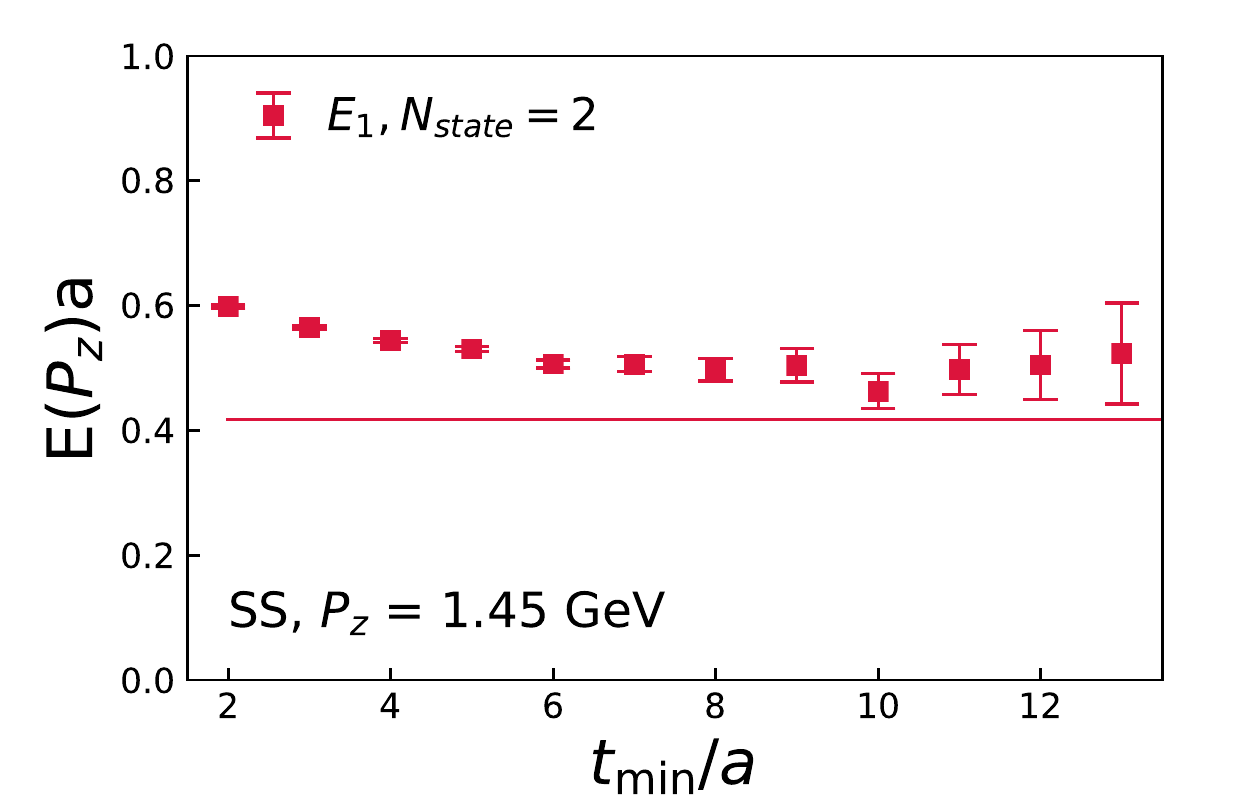}
\includegraphics[scale=0.46]{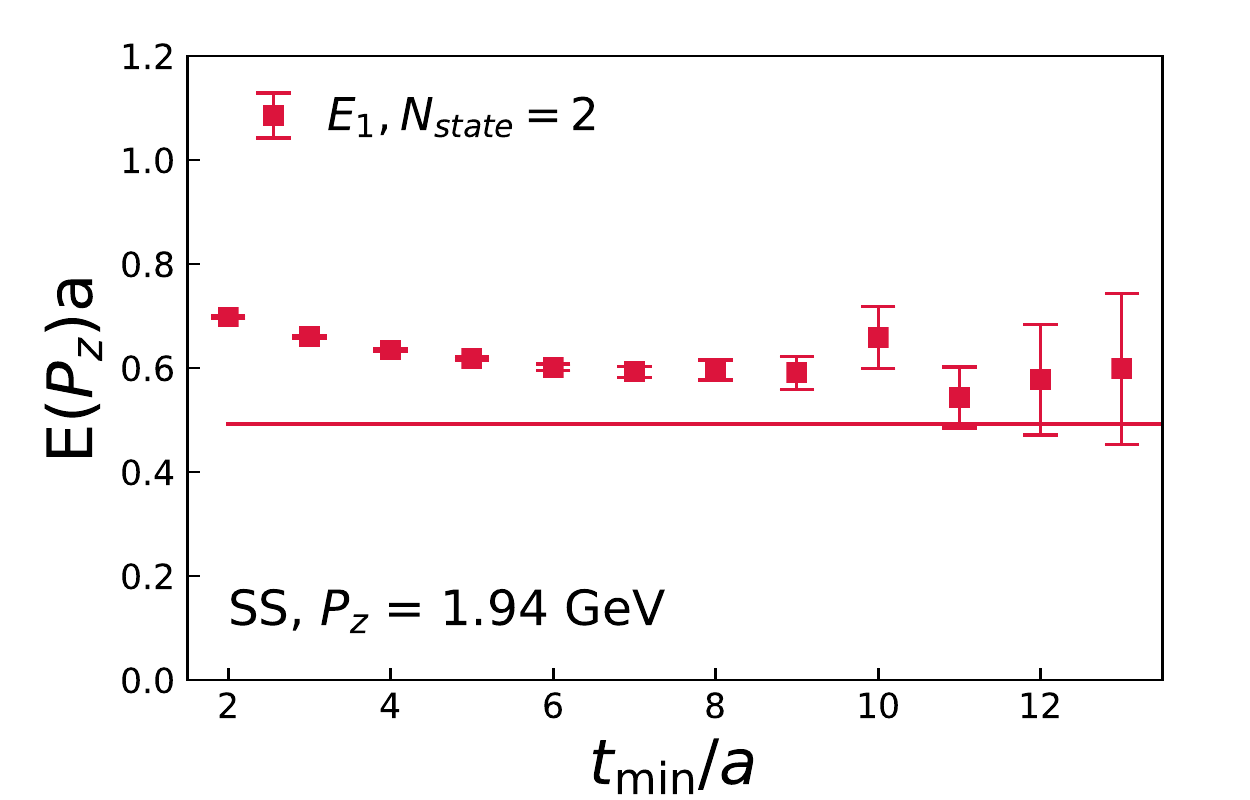}
\includegraphics[scale=0.46]{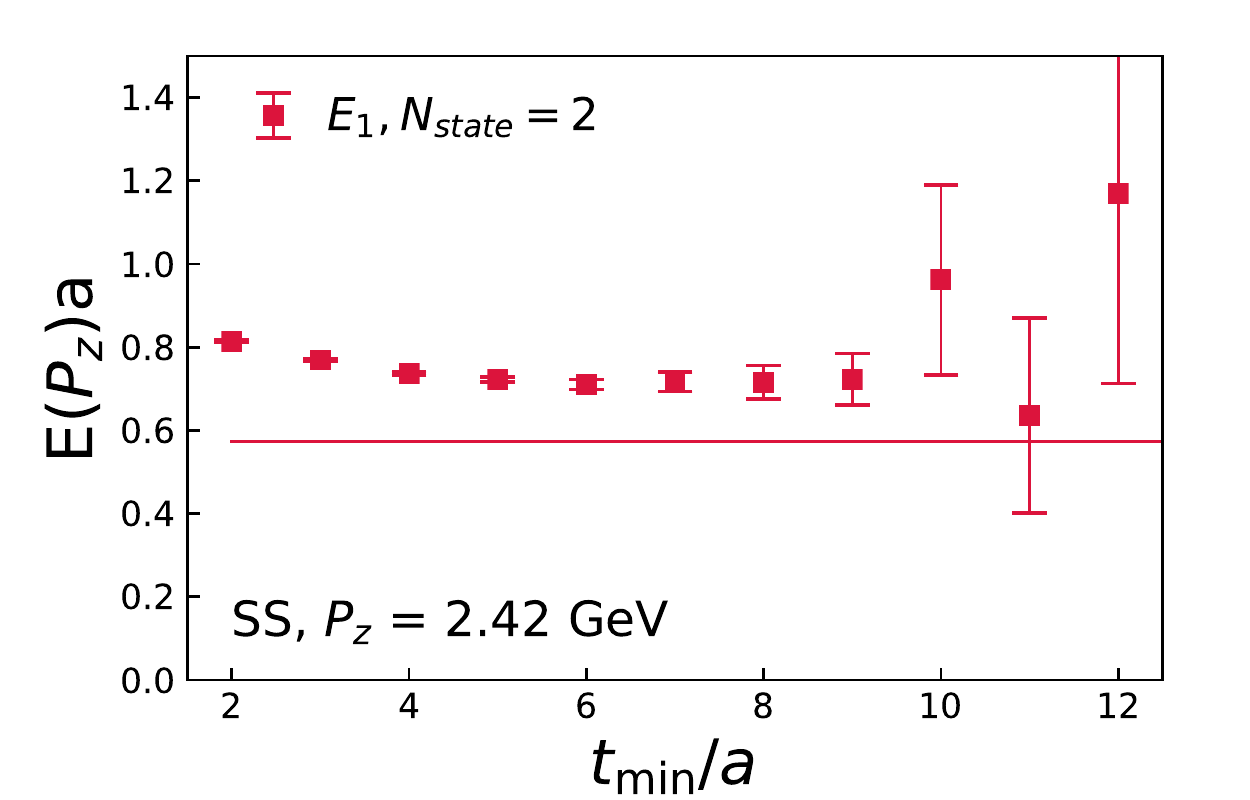}
\caption{The dependence of the first-excited state energy
$E_1(\mathbf{P})$ on the range $[t_{\rm min},32]$ used in the
two-state fits to the SS correlator is shown. The different panels
are for six different values of momentum $\mathbf{P}=(0,0,P_z)$.
For large $t_{\rm min}$, the best fit values have a tendency to
approach the dispersion values, $E_1=\sqrt{P_z^2+M_1^2}$ with
$M_1=1.5$ GeV, shown as the horizontal lines.}
\eefs{fitdetails}

It is the aim of this paper to point to the possibility of studying
the structural differences between the pion and its radial
excitation~\cite{Tanabashi:2018oca}, $\pi(1300)$. In this paper,
we will provide reasonable evidences to justify that the excited
state that we observe on the lattice shows properties of a single
particle state with similar mass to that of $\pi(1300)$, a broad
resonance state with decay-width of 200 to 600 MeV, which has been
rendered stable in the fixed finite volume of this lattice computation.
Then, we will show interesting features in the excited state bilocal
quark bilinear matrix elements and the extracted PDFs and its
moments, all under the justified hypothesis that the first excited
state on the lattice is indeed $\pi(1300)$.

\section{Details on two-point function analysis and evidences for
$\pi(1300)$ as the first-excited state}

\befs
\centering
\includegraphics[scale=0.8]{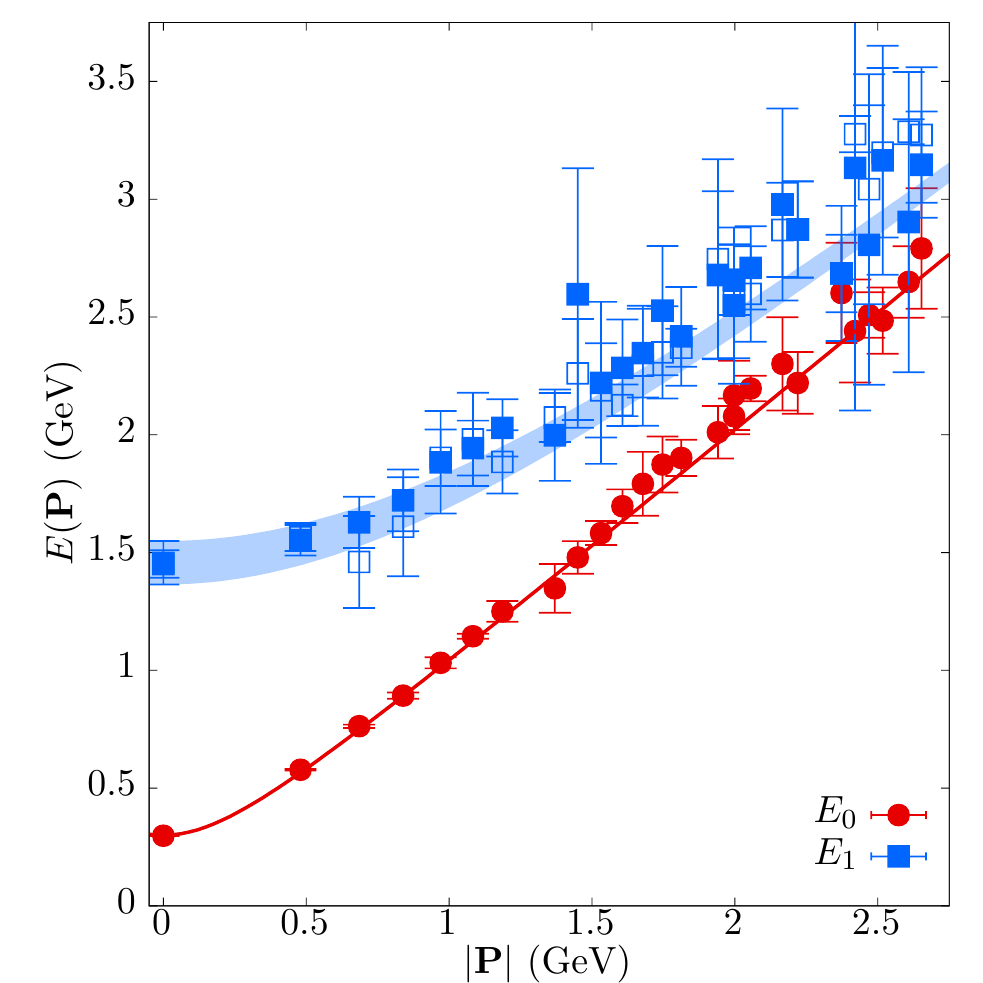}
\includegraphics[scale=0.8]{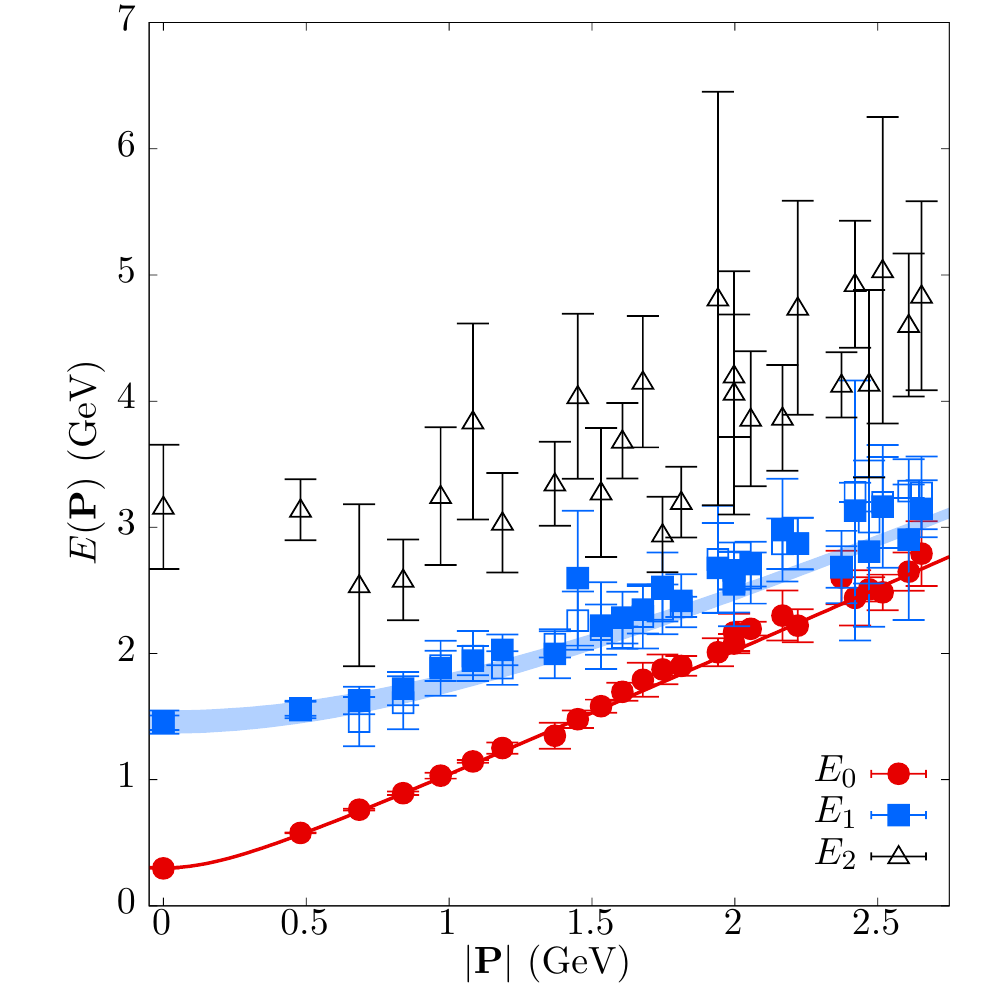}
\caption{
Observation of particle-like dispersion for ground-state and first
excited state in pion correlator.  (Left) The energies of first two
excited states as extracted from two-state and three-state fits to
the pion two-point function are shown as a function of the magnitude
of spatial momentum $|\mathbf{P}|$. The red circles are for the
ground state pion using two-state fit to the SS correlator.  The
blue symbols are for $E_1$; the filled ones correspond to estimates
from two-state fits and the open ones to the estimates from three-state
fits with priors based on the estimates from the SP correlator.
The values of $E_1$ from the two types of fits agree well with the
particle-like dispersion $\sqrt{|\mathbf{P}|^2+E_1(0)^2}$ (blue
band).  (Right) The second excited state $E_2$ from the three-state
fits are shown in addition to $E_0$ and $E_1$.
}
\eefs{piondisp}

In Refs~\cite{Gao:2020ito,Izubuchi:2019lyk}, we previously studied
the valence PDF of pion at two fine lattice spacings of 0.04 fm and
0.06 fm.  In this section, we discuss the numerical evidences in
these previous computations that the first excited state, occurring
in the spectral decompositions of the two-point and the three-point
functions, is likely to be a single particle state, and that it
corresponds to the first pion radial excitation, $\pi(1300)$. We
will do this by first showing that the excited state energy obtained
by the two- and three-state fits to the pion two-point function is
consistent with a single particle energy-momentum dispersion relation.
Then, we will notice that the mass of this state obtained from the
$P_z=0$ correlator lies close to $1.3$ GeV, the pole mass of
$\pi(1300)$, and the discrepancy is only about 200 MeV.  A source
of this discrepancy could simply be the heavier than physical pion
mass used in this work. Another source could be that the first
excited state is computed in  a fixed finite volume, and it can
differ from the pole mass of the actual resonance in the infinite
volume limit.  Below, we elaborate further.

In this work, we solely concentrate on the $a=0.04$ fm lattice
spacing ensemble used in~\cite{Gao:2020ito}, which consists of
$L_t\times L^3$ lattices with $L_t=L=64$.  We used Gaussian
smeared-source smeared-sink set-up (SS), as well as the smeared-source
point-sink set-up (SP) to determine the two-point functions of pion,
\beq
C_{\rm 2pt}(t_s) = \left\langle \pi(\mathbf{P},t_s) \pi^\dagger(\mathbf{P},0)\right\rangle.
\eeq{twopt}
In the above construction, we used momentum (boosted) quark
smearing~\cite{Bali:2016lva} to improve the signal for the boosted
hadrons.  We have discussed the details of the parameters used in
the source-sink construction, as well as our analysis methods for
the two-point function in our previous publication Ref~\cite{Gao:2020ito}.
It is worth pointing out that we were able to obtain a visible
signal for the first excited state in the $a=0.04$ fm computation,
that we will describe in the next section, because the smearing
radius of the quark sources was kept constant in lattice units
instead of in physical units; namely, for the $a=0.06$ fm ensemble
with an optimal tuning, the radius of Gaussian source was $0.312$
fm, whereas on the $a=0.04$ fm ensemble, our choice resulted in a
radius of $0.208$ fm which is smaller than the optimal one.

We analyzed the spectral content of the two-point
function through fits to the two- and three-state Ansatz; namely
\beq
C_{\rm 2pt}(t_s)=\sum_{i=0}^{N_{\rm state}-1} |A_i|^2 \left( e^{-E_i t_s}+ e^{-E_i(L_t-t_s)}\right),
\eeq{ansatz}
with $N_{\rm state}=2$ and 3 respectively. The amplitudes $A_i$ and the energies
$E_i$ are the fit parameters in this analysis.  The reason for using
two different choices of source and sink is two fold; first, the
SP correlator has a larger contribution from the excited state and
second, to check for the consistency between the energies extracted
using the two independent set of correlators.  We checked for the
robustness of the fit parameters by varying the range of source-sink
separation, $t_s\in[t_{\rm min},t_{\rm max}]$, used in the fits and
by making sure that the parameters have plateaued.  In~\cite{Gao:2020ito},
we studied only pions boosted along the $z$-direction.  For this
work, we also used pions boosted with spatial momenta
$\mathbf{P}=(P_x,P_y,P_z)$ with non-zero $P_y$ and $P_x$ for the
two-point function analysis, and obtained their ground state pion
energy $E_0$ as a function of $\mathbf{P}$ using two-state fits to
both SP and SS correlators.  We were able to isolate the ground
state energy well using a fit range shorter than $t_s\in[0.56\text{
fm}, 32a]$, whose values were consistent between both the SS and
SP correlators.  The resulting values of the ground-state $E_0(\mathbf{P})$ followed
the continuum dispersion relation,
\beq
E_0(\mathbf{P})^2=M_\pi^2+|\mathbf{P}|^2,
\eeq{dispeqn}
with $M_\pi=0.3$ GeV, to a very good accuracy even up to our largest
momentum $|\mathbf{P}|=2.42$ GeV on our fine lattice.
Having demonstrated that the actual lattice results for the ground
state satisfied our expectations about a single particle pion state,
we simply used the values of $E_0(\mathbf{P})$ from \eqn{dispeqn}
to fix the values of $E_0$ in the spectral decomposition in~\eqn{ansatz}
and determined the other free parameters; namely the amplitudes of
the ground and first excited state, and the energy of the first
excited state.

We determined the first excited state energy $E_1(\mathbf{P})$ using
(1) two-state fits to the SS and SP correlators with fixed value
taken from the dispersion relation for $E_0$, and (2) by using
three-state fits to the SS correlator with fixed $E_0$ and imposing
a prior on $E_1$ with the central value and width of the prior set
to the best fit value of $E_1$ and its error obtained from the
two-state fit to the SP correlator.  In \fgn{fitdetails}, we show
the dependence of $E_1$ on the fit range $[t_{\rm min},32a]$.  Each
panel corresponds to the six different momenta $\mathbf{P}=(0,0,P_z)$,
and for each momentum, we have shown the $t_{\rm min}$ dependence
of $E_1$ from the two-state fit.  The best fit values of $E_1$
plateau for $t_s\ge 10 a$. First, we notice that the best fit value
of $E_1(P_z=0)=1.456(92)$ GeV,  numerically lies close to the central
value of the 1.3 GeV physical mass of the pion radial excitation.
This difference of about 200 MeV between the lattice result and
physical value for the first excited state is also close to the 150
MeV difference between the mass of pion in our lattice computation
and physical pion mass.  This observation initially lead us to
identify the first excited state on the lattice with the pion radial
excitation.

In \fgn{fitdetails}, we also show the value of $E_1(\mathbf{P})$
expected from a single particle dispersion relation with a mass
$M_1 =E_1(\mathbf{P}=0)=1.5$ GeV. We find that the best fit values
of $E_1$ indeed approach the expected continuum values for non-zero
momenta.  This shows that the first excited state is likely to be
a single particle eigenstate, and not a pseudo single particle state
that effectively captures a continuum of multiparticle states.  In
the case of pion, such a possible multiparticle excited state is a
three pion state with zero angular momentum and with their total
isospin being 1.  For our ensemble, we estimate the invariant mass
of such a state to be 0.9 GeV, which is much smaller than the first
excited state we are finding.  One possibility is that the Gaussian
source we are using does not have an overlap with the three pion
state due to its vastly different delocalized spatial distribution
compared to a localized single particle state. To summarize our
evidence for observing $\pi(1300)$, we plot the energy-momentum
dispersion relation for the ground state and the first excited state
in the left panel of \fgn{piondisp}.  For $E_1$, we have shown its
estimates from the two-state fits with prior on $E_0$, and from
three-state fits with prior on $E_0$ and $E_1$ as described above
--- these are shown as the blue filled and open squares in the
figure, and they can be seen to agree well with each other.  We
find that both $E_0$ and $E_1$ agree with their respective single
particle dispersion curves. From the three-state fits with priors
on $E_0$ and $E_1$, we were able to estimate the second excited
state $E_2$, which must capture the tower of excited states above
$\pi(1300)$. In the right panel of \fgn{piondisp}, we have shown
these estimates for $E_2$ as the black triangle points, and shown
it in comparison to $E_0$ and $E_1$. We will use results on the
spectrum from the three-state fit in the further analysis of
three-point functions.

Having demonstrated that the first-excited state observed in our
computation is likely to be the first radial excitation of pion,
we will henceforth work under the assumption that this is indeed
the case, and ask for the properties of this excited state given
this justified assumption. In the rest of the paper, we will refer
to the first excited state in our lattice computation by $\pi'$,
rather than calling it as $\pi(1300)$. This is because the mass of
the first excitation on our lattice is not 1300 MeV, and for the
sake of brevity. We will also simply label the first excited state
mass $M_1$ as $M_{\pi'}$.

\section{Extraction of excited state bilocal quark bilinear matrix
element}

\befs
\centering
\includegraphics[scale=0.72]{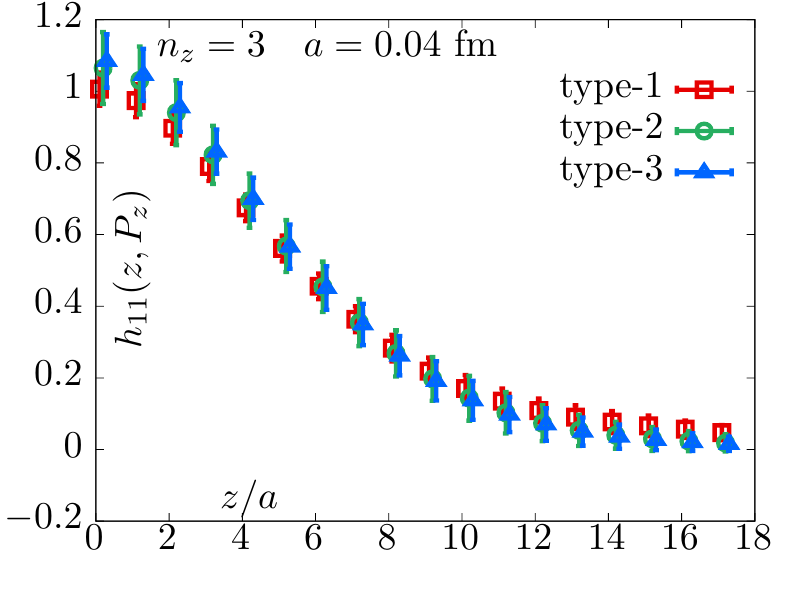}
\includegraphics[scale=0.72]{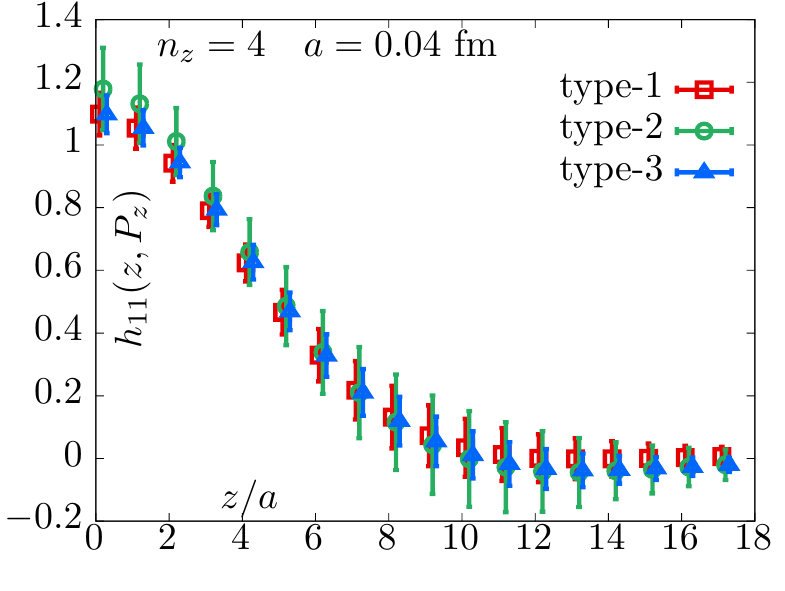}
\includegraphics[scale=0.72]{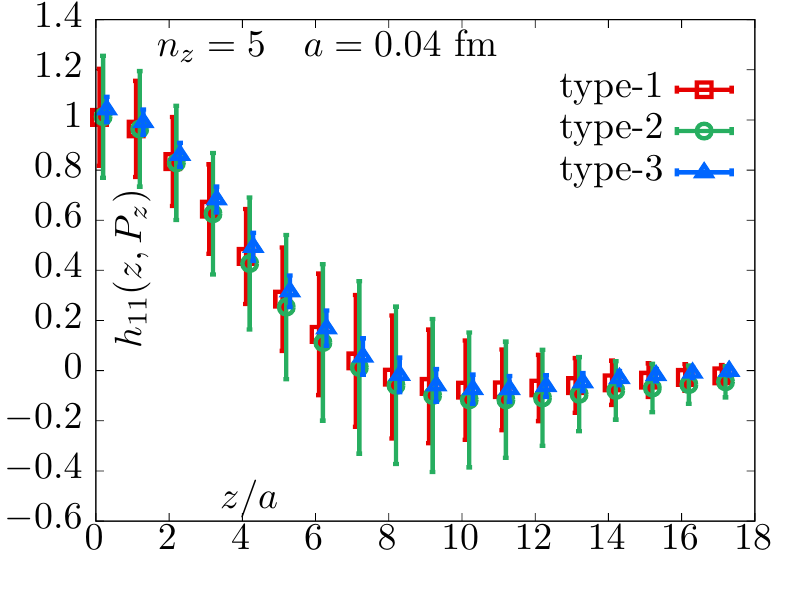}
\caption{The bare matrix element, $h_{11}(z,P_z)$, for the first
excited state of pion is shown using three different types of
three-state fits. The panels from left to right correspond to
$n_z=3,4,5$ momenta respectively.  The three types are distinguished
by the colored symbols.}
\eefs{bare}

In order to determine the required bilocal quark bilinear matrix
elements of the boosted $\pi$ and $\pi'$, we computed
the three-point function
\beq
C_{\rm 3pt}(z,\tau,t_s)=\left\langle \pi_S(\mathbf{P},t_s) O(z,\tau) \pi_S^\dagger(\mathbf{P},0)\right\rangle,
\eeq{3pt}
with both the source and sink smeared.  The bilocal operator involving
quark and antiquark separated spatially by distance $z$ is
\beqa
O(z,\tau)&=&\sum_{\mathbf{x}}\bigg{[}\overline{u}(x+{\cal L})\gamma_t W_z(x+{\cal L},x) u(x)\cr&&\ -\overline{d}(x+{\cal L})\gamma_t W_z(x+{\cal L},x) d(x) \bigg{]},
\eeqa{qpdfop}
where $x=(\tau,\mathbf{x})$ with $\tau$ being the time-slice where
the operator is inserted, and the quark-antiquark being displaced
along the $z$-direction by ${\cal L}=(0,0,0,z)$.  The bilocal
operator is made gauge-invariant by using a straight Wilson-line
$W_z$ constructed out of 1-HYP smeared gauge links.  Since we are
interested only in the parton distribution function in this paper,
we used $\mathbf{P}=(0,0,P_z)$ that is along the direction of
Wilson-line for the three-point function computations.  We used,
\beq
P_z = \frac{2\pi}{a L} n_z,
\eeq{nzval}
with $n_z=0,1,2,3,4,5$.
The spectral decomposition of the 
three-point function 
\beq
C_{\rm 3pt}(t_s,\tau;z,P_z)=\sum_{i,j} A_i^* A_j h_{ij}(z,P_z) e^{-E_i(t_s-\tau)-E_j\tau},
\eeq{3ptspec}
contains information on all the 
matrix elements between $i$-th and $j$-th states with pion quantum number
\beq
h_{ij}(z,P_z) = \langle E_i,P_z| O(z)|E_j,P_z\rangle.
\eeq{medef}
We obtained the values of the amplitudes $|A_i|$ and the energies
$E_i$ from the analyses of $C_{\rm 2pt}$.  We fixed them to the
central values from the three-state fits. One can extract the matrix
elements $h_{ij}$ by fitting the $t_s$ and $\tau$ dependence of
$C_{\rm 3pt}$ data to the spectral decomposition above, with $h_{ij}$
being the unknown fit parameters.  In practice, for the cross-terms
such as $A_0^* A_1 h_{01}$, we simply treated the real part of such
whole factors together as the fit parameters, whereas for the
diagonal terms only the magnitudes $|A_i|$ enter, and therefore,
we were able to resolve the diagonal matrix elements $h_{ii}$ without
any phase ambiguity.

We implemented this analysis by first forming the standard ratio 
\beq
R(t_s,\tau;z,P_z)\equiv \frac{C_{\rm 3pt}(t_s,\tau;z, P_z)}{C_{\rm 2pt}(t_s;P_z)},
\eeq{ratiodef}
so that the leading term in this ratio as $t_s\to\infty$ is the
ground state matrix element $h_{00}$.  In \cite{Gao:2020ito}, we
presented detailed analysis of the ratio $R$ using both the two-state
and three-state fits. In that work, we found that a simple two-state
fit was enough to obtain $h_{00}$ which was consistent with a more
elaborate three-state fit as well as with the summation method. On
the one hand, a simple two-state fit is not justified here, as we
are interested in the first excited state, and therefore, at least
one more state other than the first excited state should be included
in the analysis.  On the other hand, a full three-state analysis
involving 9 independent fit parameters will make the determinations
of $h_{11}$ noisy.  Therefore, we experimented with variations of
the three-state fit by reducing the number of parameters in the fit
and by imposing prior on the ground state matrix element $h_{00}$
from the two-state fit. We first performed the full three-state fit
with 9 parameters, which we call as the fit of type-1. Then, in a
fit of type-2, we imposed a prior on $h_{00}$ keeping all other fit
parameters of the full state fit to be free; for the prior and its
width we took the value of $h_{00}$ and its statistical error from
the two-state analysis of the three-point function.  In a fit of
type-3, in addition to imposing the prior on $h_{00}$, we also
assumed that we can ignore the second-excited state matrix element
$h_{22}$, thereby reducing the fit parameters to 8 (or effectively
7, due to the prior). In all the three Ansatze, we kept all the
matrix elements which involved the first excited state.

In \fgn{bare}, we show $\pi'$ matrix element, $h_{11}$, as a function
of $z$ at the three largest momenta. The different colored symbols
are the extrapolations using the above three types of three-state
Ansatz. For $n_z=3$, the type-1, nine-parameter three-state fit
actually performs better than when constraints are imposed. However,
this is not true at the higher $n_z=4,5$ momenta, which are crucial
to ensure that the momenta are larger than the $\pi'$ mass. For
$n_z=3$, the type-2 fit results in noisier estimates of $h_{11}$
compared to type-1, whereas the type-3 fit results are consistent
with type-1 results with a slight reduction in errors.  Therefore,
we find that the type-3 Ansatz leads to a reasonable reduction in
the statistical errors with only the assumption that $h_{22}$ matrix
element can be ignored.  In fact, from the unconstrained type-1
fits, we found that the resulting values for $h_{22}$ were consistent
with zero and it was merely making the results noisier. Therefore,
the usage of type-3 Ansatz to obtain better estimates of $h_{11}$
seems to be justified.  We tried reducing the number of parameters
further by ignoring the cross-terms $h_{12}$ and $h_{21}$, but it
resulted in unreasonably ultra-precise estimations of $h_{11}$,
showing that such constraints rule out most of the parameter space
--- it would have been a positive outcome if there was a strong
theoretical underpinning to ignoring the cross-terms, but in the
absence of such a justification, we avoided using such stricter
constraints.  From the fit results for $n_z=5$ shown in the rightmost
panel of \fgn{bare}, the usage of type-3 fit renders $h_{11}$ at
this momentum usable. In the analysis of PDF that follows, we will
use the values of $h_{11}$ obtained using type-3 Ansatz for the
extrapolations, and we will also show results from the type-1 fits
to contrast it against.

The bilocal operator $O$ needs to be multiplicatively
renormalized~\cite{Ji:2017oey,Ishikawa:2017faj,Green:2017xeu}.  The
details pertaining to renormalization as applied to our computations
are described in detail in~\cite{Izubuchi:2019lyk,Gao:2020ito}. One
possibility is to determine the renormalization factors
$Z_{\gamma_t\gamma_t}(z,P^R)$ in the RI-MOM
scheme~\cite{Stewart:2017tvs,Chen:2017mzz,Alexandrou:2010me} using
off-shell quarks at momentum $P^R=(P_z^R,P_\perp^R)$,
\beq
h^R_{\pi'\pi'}(z,P_z,P^R) = \frac{Z_{\gamma_t\gamma_t}(z,P^R) h_{11}(z,P_z)}{Z_{\gamma_t\gamma_t}(0,P^R) h_{11}(0,P_z)},
\eeq{rime}
In addition to the multiplicatively renormalizing the operator, the
ratio with the corresponding matrix element at $z=0$, helps reduce
lattice corrections and any overall systematical corrections, so that
the expectation value of the isospin charge of pion is 1 by
construction at all momenta.  Another possibility is to form
renormalization group invariant
ratios~\cite{Orginos:2017kos,Izubuchi:2018srq,Fan:2020nzz,Gao:2020ito}
between the bare matrix elements at two different momenta,
\beq
{\cal M}_{\pi'\pi'}(z,P_z,P_z^0) = \left(\frac{h_{11}(z,P_z)}{h_{11}(z,P_z^0)}\right) \left(\frac{h_{11}(0,P^0_z)}{h_{11}(0,P_z)}\right).
\eeq{ratme}
In the above ratio, the UV divergence of the operator is exactly
canceled between the two bare matrix elements. Similar to an improved
version of the RI-MOM scheme we defined in \eqn{rime}, the double
ratio at non-zero $z$ and $z=0$ matrix elements in the above equation
ensures that the isospin charge is normalized to 1. Since the UV
divergence does not depend on the external states, the two matrix
elements in the ratio need not be for the same hadron. Therefore,
we also construct the following ratio using the ground state pion
matrix element as
\beq
{\cal M}_{\pi'\pi}(z,P_z,P_z^0) = \left(\frac{h_{11}(z,P_z)}{h_{00}(z,P_z^0)}\right) \left(\frac{h_{00}(0,P_z^0)}{h_{11}(0,P_z)}\right).
\eeq{ratme2}
For the above ratio, we take our determination of $h_{00}(z,P_z^0)$
from~\cite{Gao:2020ito}.  In the next section, we will discuss the
relation of the above matrix elements to the PDF via the one-loop
leading-twist perturbative matching.

Before performing any double ratio, we can use the $z=0$ renormalized
matrix element to perform a simple cross-check.  The pion source
$\pi(P_z,t_s)$ can excite only one unit of the isospin charge, and
hence each of the states that occurs in the spectral decomposition
of the pion two-point function will carry unit isospin (up to terms
due to wrap-around effects, which are negligible for heavy excited
states).  Therefore, measuring the isospin of our first excited
state before imposing any normalization condition serves a cross-check
of the excited state extrapolations.  In \fgn{z0mat}, we show
$Z_{\gamma_t\gamma_t}(z=0,P^R)h_{11}(z=0,P_z)$ as a function of
$P_z$ after renormalization in RI-MOM scheme. It is the isospin
charge modulo the quark wavefunction renormalization which is nearly
1 at this lattice spacing~\cite{Izubuchi:2019lyk}. At $P_z=0$, our
ground-state matrix element determination suffers from 2\% lattice
periodicity effects~\cite{Gao:2020ito}, that in turn affects all
the fitted parameters in the three-state fit, particularly resulting
in a value of $Z_{\gamma_t\gamma_t}h_{11}$ slightly larger than 1.
At all other non-zero $P_z$, the extracted isospin of the first
excited state is consistent with 1, lending more confidence in the
reliability of our extrapolations.
\bef
\centering
\includegraphics[scale=0.8]{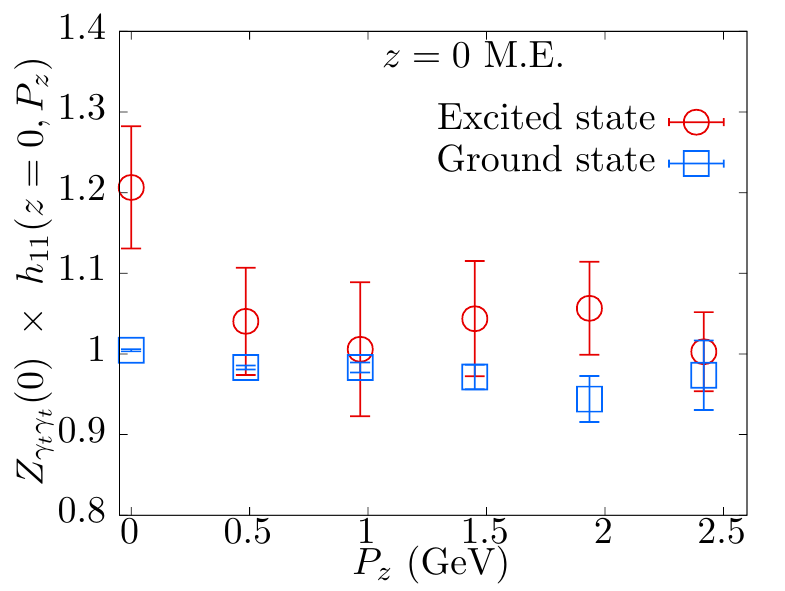}
\caption{The RI-MOM renormalized local matrix element (mod $Z_q\approx
1$) at $z=0$ with $(P^R_z,P^R_\perp)=(1.93,3.34)$ GeV is shown as
a function of $P_z$ for pion (blue) and $\pi'$
(red) as obtained from the three-state type-3 fit. The wrap-around effect
present in the three-point function at $P_z=0$ is not accounted for in the plot.}
\eef{z0mat}

\section{Comparison of the PDFs of $\pi$ and $\pi'$}

\befs
\centering

\includegraphics[scale=0.65]{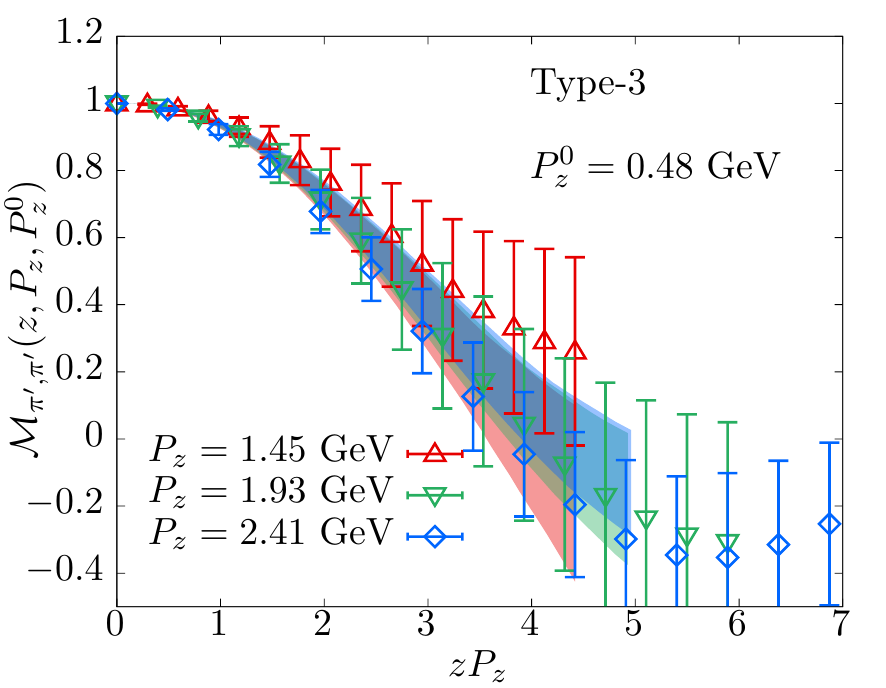}
\includegraphics[scale=0.65]{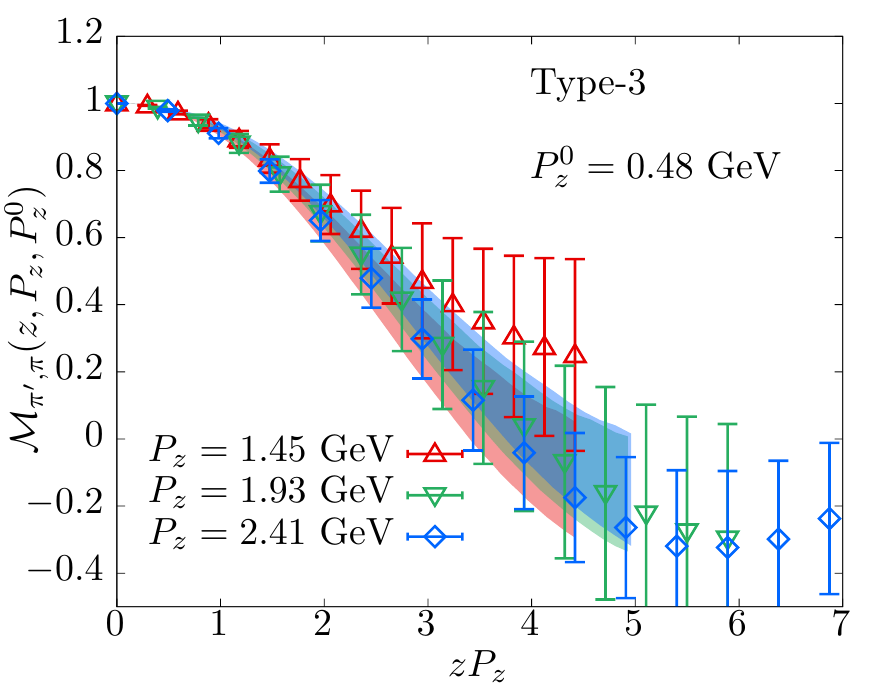}
\includegraphics[scale=0.65]{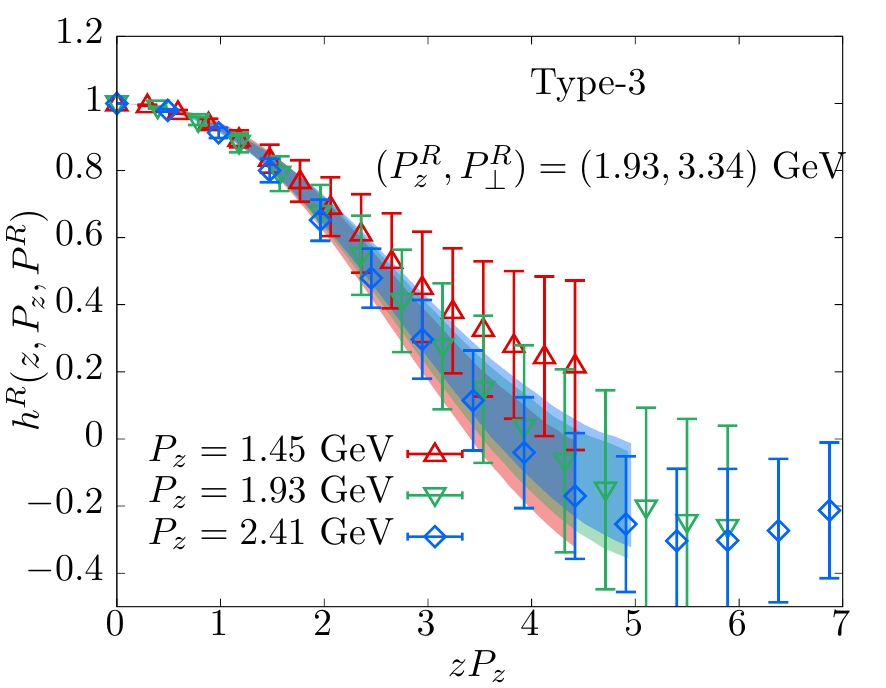}

\includegraphics[scale=0.65]{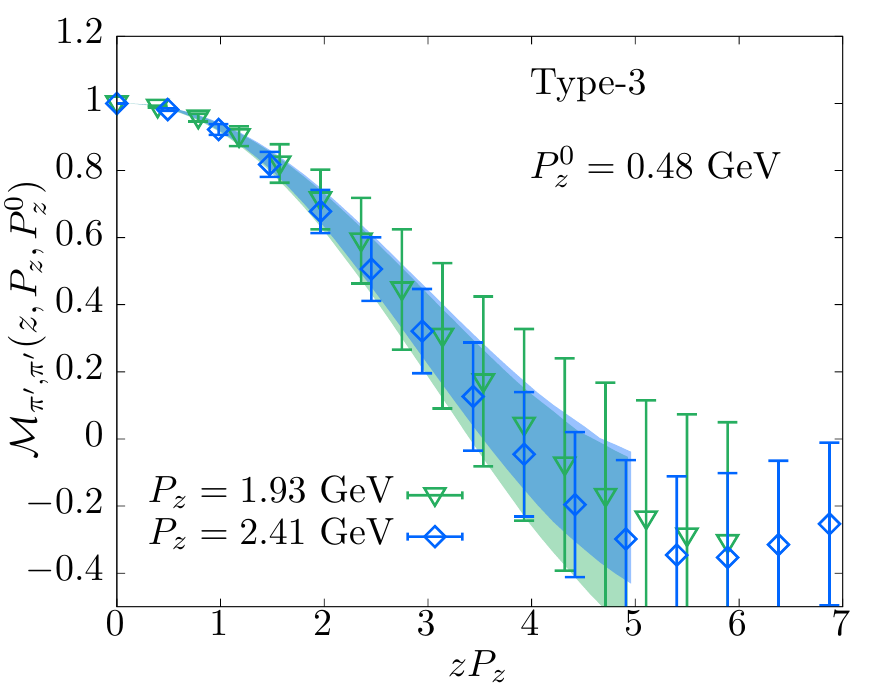}
\includegraphics[scale=0.65]{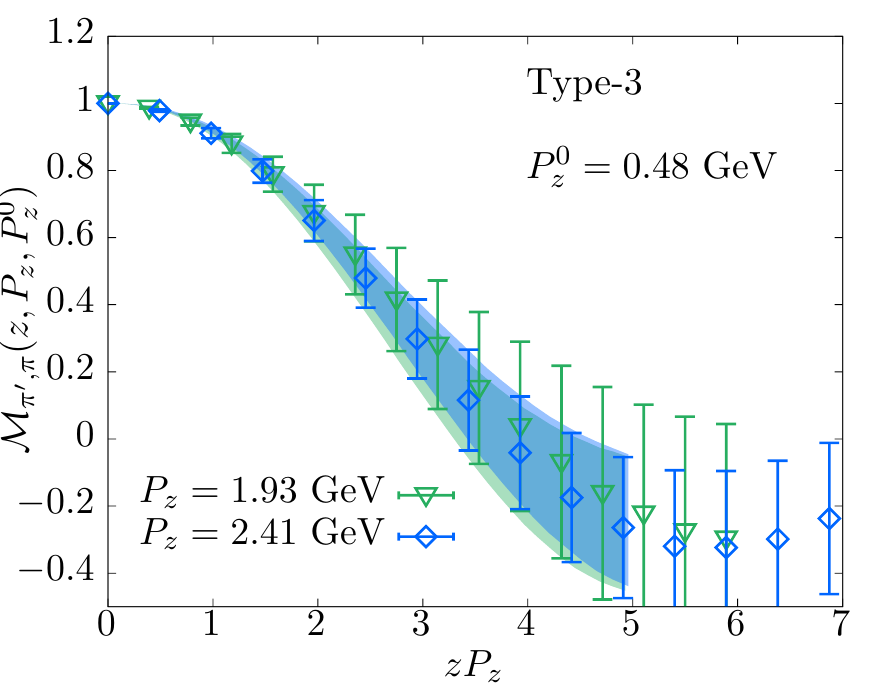}
\includegraphics[scale=0.65]{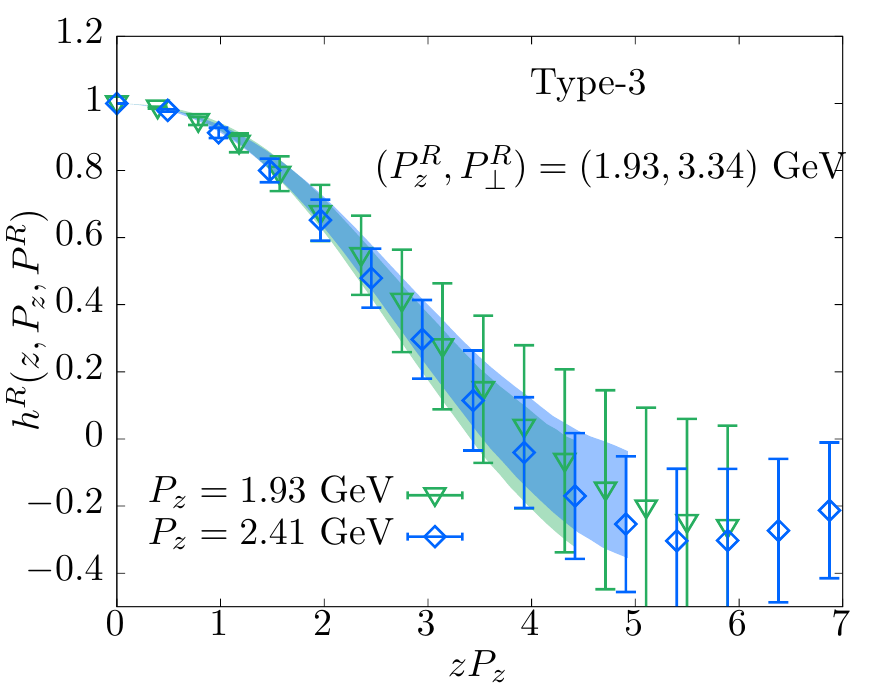}

\caption{
The excited state matrix elements renormalized in three different
schemes (ratios ${\cal M}_{\pi'\pi'}$, ${\cal M}_{\pi'\pi}$ and RI-MOM $h^R$
from left to right) are shown as a function of $zP_z$. The data
points of the same momenta are shown using same colored symbols.
The top panels show largest three momenta and the bottom one includes
only the largest two.  The bands are the fits using the respective
leading-twist OPEs to data assuming a two-parameter functional form
of the PDF.  }

\eefs{itdplot}

We used our estimates of $h_{11}$ from the type-1 and type-3 fits
to obtain the PDF of $\pi'$.  For this, we used the
twist-2 OPE expressions~\cite{Izubuchi:2018srq} corresponding to
the renormalized matrix elements described above. For the RI-MOM
matrix element at renormalization scale $P^R$, the twist-2 expression
is
\beq
h^R_{\pi'\pi'}(z,P_z,P^R) = 1+\sum c_n^{RI}(\mu,P^R,z^2) \langle x^n\rangle_{\pi'} \frac{(-iP_z z)^n}{n!},
\eeq{rimomexp}
where the sum above runs over only the even values of $n$ for the
valence PDF of the pion and its excitations due to the isospin
symmetry. The 1-loop expression for the RI-MOM Wilson coefficients
is given in~\cite{Gao:2020ito} using results
in~\cite{Constantinou:2017sej,Zhao:2018fyu}.  The terms  $\langle
x^n\rangle_{\pi'}(\mu) = \int_0^1 x^n f_v(x,\mu) dx $ are the moments~\footnote{The
nomenclature followed in this paper is such that $\langle x^n\rangle$
is the $n$-th moment.} of the valence PDF $f_v(x,\mu)$ of the first
excited state in the $\overline{\rm MS}$ scheme at factorization
scale $\mu$. We will consistently use $\mu=3.2$ GeV for all the
determinations in this paper. The twist-2 OPE expression for the
ratio scheme~\cite{Izubuchi:2018srq} is
\beq
{\cal M}_{\pi'\pi'}(z,P_z,P_z^0) = \frac{1+\sum c_n(\mu^2 z^2) \langle x^n\rangle_{\pi'} \frac{(-iP_z z)^n}{n!}}{ 1+\sum c_n(\mu^2 z^2) \langle x^n\rangle_{\pi'} \frac{(-iP^0_z z)^n}{n!}},
\eeq{ratexp}
using the expressions for the Wilson coefficients $c_n$ given
in~\cite{Izubuchi:2018srq,Radyushkin:2017lvu}. Here, we also present results using a variant of the
ratio scheme ${\cal M}_{\pi'\pi}$ described in the last section,
and it has the leading twist expression,
\beq
{\cal M}_{\pi'\pi}(z,P_z,P_z^0) = \frac{1+\sum c_n(\mu^2 z^2) \langle x^n\rangle_{\pi'} \frac{(-iP_z z)^n}{n!}}{ 1+\sum c_n(\mu^2 z^2) \langle x^n\rangle_{\pi} \frac{(-iP^0_z z)^n}{n!}},
\eeq{ratexp2}
where the moments $\langle x^n\rangle_{\pi}$ are those of the ground
state pion. We take their values from our analysis of pion on the
same ensemble presented in~\cite{Gao:2020ito}. Since the mass of
$\pi'$ is about 1.5 GeV, we took care of target mass correction at
leading twist by replacing $(P_z z)^n \to (P_z z)^n \sum_{k=0}^{n/2}
\frac{(n-k)!}{k!(n-2k)!}\left(\frac{M_{\pi'}^2}{4P_z^2}\right)^k$
in the above expressions~\cite{Chen:2016utp,Radyushkin:2017ffo}.
We work under the assumption that any target mass correction that
can occur at higher twist are negligible. In order to justify this
further, we eventually used only the matrix elements at the two
highest momenta corresponding to $P_z=1.93$ and 2.42 GeV as we
discuss below.

We performed two kinds of analysis. In a model independent analysis,
we fitted the renormalized matrix elements spanning a range of
$P_z>P_z^0$ and $z\in[z_{\rm min}, z_{\rm max}]$ using their
respective leading twist expressions above, with the even moments
$\langle x^n\rangle_{\pi'}$ as the independent fit parameters.  In
the second kind of model dependent analysis, we assumed a two-parameter
functional form of valence excited state PDF,
\beq
f_v(x) = {\cal N} x^{\alpha}(1-x)^\beta,
\eeq{pdfansatz}
and fitted the resulting moments (which are functions of $\alpha$
and $\beta$) to best describe the $z$ and $zP_z$ dependences of the
data.  This enabled us to reconstruct the $x$-dependent PDF. Since
the data for the excited state is noisy, we could not improve the above
parametrization by adding additional small-$x$ terms, as we did for
the pion in~\cite{Gao:2020ito}. In the future, one needs to perform 
a similar analysis with multiple functional forms of the PDF Ansatz
to quantify the amount of systematic error.

We first describe our reconstruction of the PDF using the two-parameter
functional form.  In \fgn{itdplot}, we put together the renormalized
bilocal matrix elements~\footnote{ The quantity $zP_z$ has also
been referred to as the Ioffe-time~\cite{Braun:1994jq}, and the
bilocal matrix element is also referred to as the Ioffe-time
Distribution~\cite{Radyushkin:2017cyf}.  In the lack of a short-distance
limit or infinite momentum limit, the matrix element is common and
exactly the same for both LaMET as well as the short-distance
factorization used in pseudo-PDF approach. Therefore, we refer to
the renormalized matrix elements as simply bilocal matrix elements,
without any ambiguity.  } at different fixed momenta and show them
as a function of $zP_z$.  The left, middle and the right panels are
in the two ratio schemes, ${\cal M}_{\pi'\pi'}$ and ${\cal
M}_{\pi'\pi}$, and in the RI-MOM scheme with $(P^R_z,P^R_\perp)=(1.93,
3.34)$ GeV respectively. We have used the first non-zero momentum,
$P_z^0=0.48$ GeV as the reference momentum to construct the ratios,
which is slightly above $\Lambda_{\rm QCD}$ and also contributes
minimally to the statistical noise. In the top panels, the data
from the three highest momenta are shown, whereas in the bottom
panels, only the two highest momenta, $P_z=1.93$ and 2.41 GeV, which
are larger than the excited state mass of 1.5 GeV are shown. The
bands are the expectations based on the best fits using the
two-parameter PDF Ansatz; the bands are colored in the same manner
as the corresponding data points at different momenta.  For the
cases shown in the top panel, we performed the fits using all the
three momenta shown, and over a range of quark-antiquark separation
$z\in[2a,0.6 {\ \rm fm}]$. Given the noisy data compared to that
of the ground state pion, we could not perform an {\sl ideal}
analysis, where one would want to keep range of $z$ even smaller
than what is used here. We skipped $z=a$ to avoid the $(P_z a)^2$
lattice correction~\cite{Gao:2020ito}. Overall, the fits can be
seen to perform well regardless of the momenta included in the
analysis. However, upon a close inspection of the analyses in the
top-panel, we find that the evolution of the data with $P_z$ at
different fixed $z P_z$ has opposite trends between the data and
the fits; namely, the central values of the data have a decreasing
tendency from $P_z=1.45$ GeV to 2.41 GeV (albeit well within errors),
whereas the fitted bands have the opposite behavior. This indicates
the presence of possible higher twist corrections when matrix
elements at momentum $P_z=1.45$ GeV, which is comparable to the
mass of the excited state, is included in the analysis.  On the
other hand, in the lower-panel, the data at the two highest momenta
are compatible with each other and the fitted bands are also seen
to be describing the data well. Therefore, to be cautious, we will
simply include the data at the two highest momenta in the analysis
henceforth.

\befs
\centering

\includegraphics[scale=1.1]{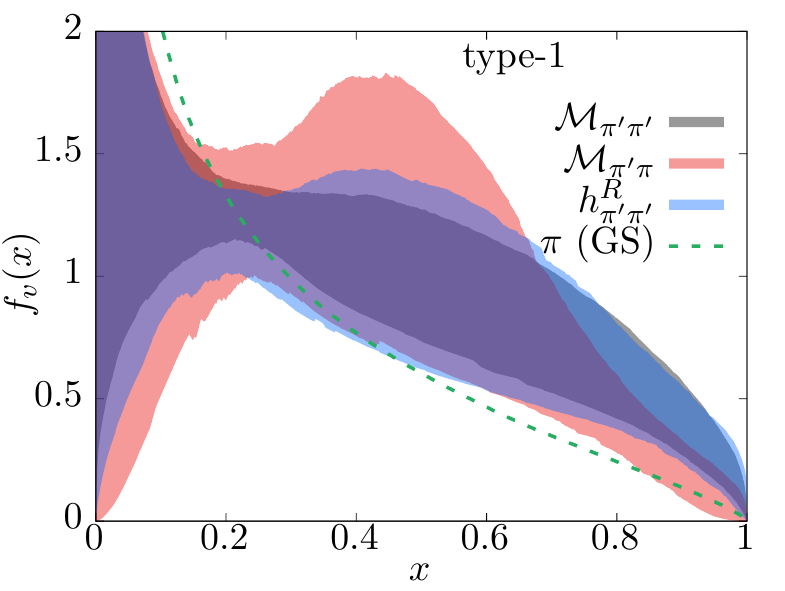}
\includegraphics[scale=1.1]{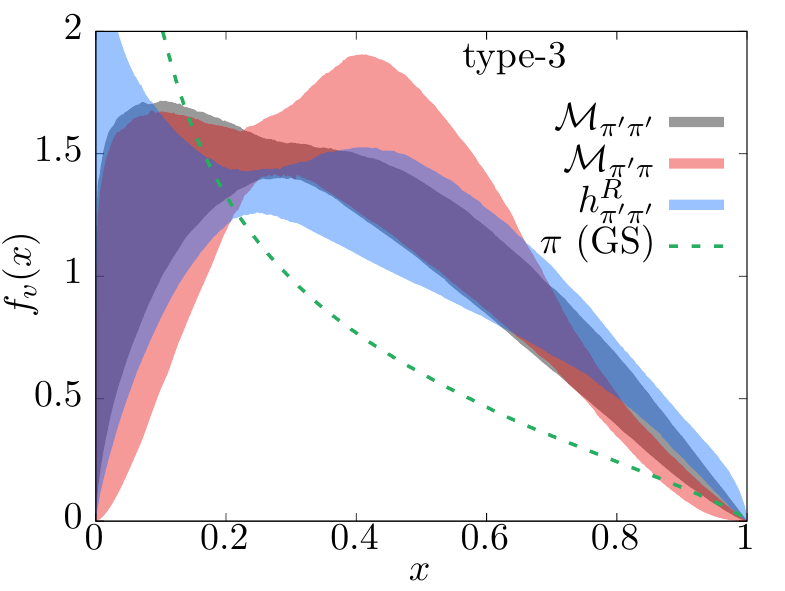}
\caption{
    The valence parton distribution function $f_v(x,\mu)$ of $\pi'$
    at $\mu=3.2$ GeV, as reconstructed from real space matrix
    elements using fits to simple two-parameter Ansatz $f_v(x,\mu)={\cal
    N} x^{\alpha} (1-x)^\beta$.  The results for the valence PDF
    obtained using the bilocal matrix elements in three different
    renormalization schemes are shown using the different colored
    bands. The results using type-1 and type-3 matrix elements are
    shown on the left and right panels respectively.  The valence
    PDF of pion as determined using the same ensemble and analysis
    methods as used for $\pi'$ is shown as the dashed curve for
    comparison.
}
\eefs{pdfplot}

In \fgn{pdfplot}, we show the $x$-dependent valence PDF of the
excited state, $f_v(x)$, that is reconstructed based on the
two-parameter Ansatz fits in the real-space shown as bands in the
bottom panels of \fgn{itdplot}.  The right panel of \fgn{pdfplot} is based on fits
to the matrix elements obtained using type-3 extrapolation, whereas
the left one is using the type-1 extrapolation.  We have
compared the PDF determinations as obtained from the fits to the
matrix elements in the three different renormalization schemes. For
comparison, the central value of the ground state pion PDF from the
same ensemble~\cite{Gao:2020ito} is shown as the dashed green line.
First, the usage of type-1 extrapolated matrix elements results in
very noisy PDF that cannot be used to find any hints of structural
differences; within the large errors, the excited pion PDF is
consistent with the ground state PDF. On the other hand, the usage
of type-3 extrapolated matrix element does result in better determined
PDFs. Therefore, let us focus on the right panel of \fgn{pdfplot}.
The consistency amongst the estimates from different renormalization
schemes, which differ also in their matching formulas, is reassuring. 
Using ${\cal M}_{\pi'\pi'}$, we found the PDF is parametrized by 
$\left\{\alpha,\beta\right\}=\left\{0.4(3), 1.1(2)\right\}$.
It is very clear that the PDF of the radial excitation is different
from the ground state --- the excited state PDF is consistently
above the pion PDF starting from an intermediate $x\approx 0.3$ to
large-$x$. There is a tendency in the excited state PDF to vanish
at small-$x$, but it is not conclusive given the errors and
also due to possibly large higher-twist effects contaminating the
small-$x$ regime.  Thus, the overall trend seems to be that the
valence PDF of the radial excitation is shifted towards larger-$x$
compared to the ground state valence PDF. This points to smaller
momentum fraction being carried by gluons and sea quarks in the
radially excited state compared to the pion. 

In \fgn{momplot}, we compare the first four valence PDF moments of
the radial excitation obtained using the model-dependent and
model-independent analyses. The results for $\pi'$ from various
fitting procedures are shown on the left part of the plot, and the
values for the pion, taken from our previous work on the same
ensemble~\cite{Gao:2020ito}, are shown on the right part of the
figure.  For the model-independent fits, we used the first three
even moments $\langle x^2\rangle, \langle x^4\rangle$ and $\langle
x^6\rangle$ themselves as the fit parameters. We also imposed the
inequality conditions between the valence moments as discussed
in~\cite{Gao:2020ito}.  Similar to the PDF Ansatz fits, we present
the results of the fits over a $z$-range of $[2a,0.6 {\rm fm}]$ in
\fgn{momplot}; the model-independent fits to type-1 and type-3
matrix elements are labeled as A and B in \fgn{momplot}. Since we
cannot determine the odd moments directly by this model-independent
procedure, only the results for $\langle x^2\rangle$ and $\langle
x^4\rangle$ are shown for them.  The results for the moments as
inferred from the two-parameter fits, using the relation $\langle
x^n\rangle=\int_0^1 x^n f_v(x) dx$, are also shown for $\pi'$ in
\fgn{momplot}; the results from fits to type-1 and type-3 ${\cal
M}_{\pi'\pi'}$ are labeled as C and D, whereas the ones from fits
to type-3 ${\cal M}_{\pi'\pi}$ matrix element are labeled as E. It
is comforting that the PDF Ansatz fits result in values of the even
moments that are consistent with those from the model-independent
fits.  This also gives us the confidence in the indirect determination
of the odd moments via this procedure.  To justify this indirect
method, taking the case of pion where the phenomenological values
of the odd moments are known, in~\cite{Gao:2020ito}, we found that
similar analysis via fits to PDF Ansatze resulted in values of the
odd moments that agreed reasonably well with the phenomenological
values.
\bef
\centering
\includegraphics[scale=1.0]{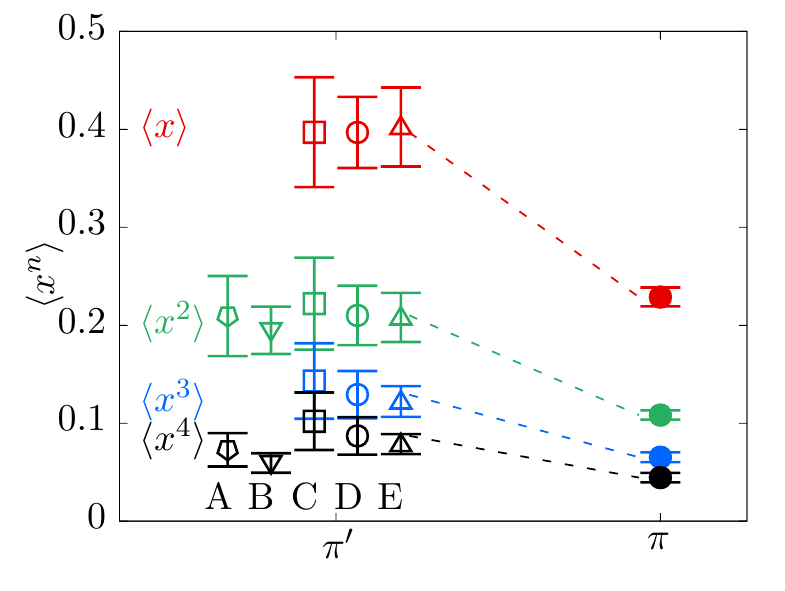}
\caption{
    The lowest four valence PDF moments of the pion radial excitation
    $\pi'$ is compared with those of the ground state pion $\pi$.
    For the radial excitation, both the even and odd moments extracted
    from the two-parameter PDF Ansatz fit are shown along with
    direct model independent estimates of even moments: (A)
    model-independent fit to type-1 ${\cal M}_{\pi'\pi'}$.  (B)
    model-independent fit to type-3 ${\cal M}_{\pi'\pi'}$.  (C) PDF
    Ansatz fit to type-1 ${\cal M}_{\pi'\pi'}$.  (D) PDF Ansatz fit
    to type-3 ${\cal M}_{\pi'\pi'}$.  (E) PDF Ansatz fit to type-3
    ${\cal M}_{\pi'\pi}$.  The dashed line connect the central
    values of $\pi'$ moments to that of $\pi$, to aid the eye.
}
\eef{momplot}

It is at once striking that the moments of $\pi'$ are larger than
that of the pion, especially in the case of the lowest two-moments
$\langle x\rangle$ and $\langle x^2\rangle$.  Quantitatively, by
taking the values of $\left\{\langle x\rangle,\langle x^2\rangle\right\}$
from the method ``D" in \fgn{momplot}, we see that they are
$\left\{0.40(4), 0.21(3)\right\}$ for $\pi'$, which is to be compared
with $\left\{0.2289(96), 0.1083(47)\right\}$ for the pion.  This
is the reason we observed the valence PDF of $\pi'$ to be above
that of the pion at higher values of $x$. Therefore, at a scale of
$3.2$ GeV, only 20\% of the $\pi'$ momentum fraction comes from
gluons and sea quarks, which forms a larger 56\% component for the
ground state pion. Thus, within the two-parameter PDF Ansatz analysis,
it appears that the valence quarks carry almost twice the momentum
fraction in the radial excitation of the pion compared to its ground
state.  This link could simply be a correlation or perhaps be causal,
which needs to be investigated using simpler models.

\section{Conclusions and outlook}

In this work, we presented a proof-of-principle computation of the
first excited state of the pion determined in a fixed finite volume
and at a fixed fine lattice spacing. We argued that the first excited
state is most likely to be a single particle state since its energy
satisfies a single particle dispersion relation. Also, the mass of
the state compares well with the central value of the experimentally
observed radial excitation, which is however a resonance in the
infinite volume limit. Given the observations, we hypothesized that
the first excitation on our lattice is that of the pion radial
excitation, $\pi(1300)$.  With a reasonable reduction in the number
of unknown parameters in the three-state fits to the three-point
function involving the bilocal quark bilinear operator, we were
able to extract the boosted $\pi(1300)$ matrix elements.  We performed
a model-independent analysis to obtain the even valence PDF moments,
and used model-dependent PDF Ansatz fits to reconstruct the
$x$-dependent valence PDF at a scale of $\mu=3.2$ GeV. We found
evidences that in indicate that (1) the valence PDF of $\pi(1300)$
consistently lies above that of pion for intermediate and large-$x$
regions, thereby indirectly, implying a reduced role of gluons and
sea quarks in the excited state.  (2) Quantitatively, the lower
moments of $\pi(1300)$ were about twice larger than that of $\pi$.

The present work was meant only as a pilot study towards understanding
how the ground and excited states of hadrons differ. Therefore,
this study can be made more rigorous in at least three major ways --- 
(1) One could either render the radial excitation to be stable
single-particle state by using a larger unphysical pion mass
(e.g.,~\cite{Chai:2020nxw}), or one needs to perform a dedicated
finite-size scaling study of the excited state PDF in order to make
connection with the actual resonance state in the thermodynamic
limit. (2) Usage of larger operator basis in two-point functions
that will lead to a more sophisticated spectroscopy of pion correlators
leading to a more convincing determination of the first excited
state as well as its quantum numbers.  (3) Incorporating similar
techniques for the three-point function for a reliable determination
of the excited state matrix element without involving any reduction
in number of fit parameters as done here.

As a concluding remark, in the absence of a microscopic theory of the transition from pion
to its radial excitation, we propose the following momentum
differential as a useful quantity.  To motivate the quantity, one
can consider a process, such as $\pi'\to \pi+({\pi \pi})_{\rm
S-wave}$, for a special instance with both $\pi'$ and $\pi$ after
the transition are at rest in the lab frame, and the difference in
their masses carried by other product states.  In such an artificially
constructed experimental outcome, one could ask how the change,
$\Delta P^+ = (M_{\pi'}-M_\pi)/\sqrt{2}$, compares to the change
in the average momentum $\Delta \langle k^+\rangle =(M_{\pi'}\times
2\langle x \rangle_{\pi'} - M_{\pi}\times 2\langle x
\rangle_{\pi})/\sqrt{2} $ of the two valence partons in $\pi'$ and
$\pi$.  This motivates the construction of the Lorentz invariant
ratio,
\beq
\zeta = \frac{2M_{\pi'} \langle x \rangle_{\pi'} - 2M_{\pi} \langle x \rangle_{\pi}}{M_{\pi'}-M_\pi},
\eeq{exfrac}
as a measure to correlate the structural changes to the
differences in the masses.  Using $\Delta M =1.2$ GeV, we find that
the fraction $\zeta$ ranges from 0.78 to 0.99 given the variations
within 1-$\sigma$ errors on the first moments we discussed above.
Even if we discount a 2-$\sigma$ variation, the fraction is at least
0.68.  Even such a simple-minded modeling of the excitation tells us
that the changes to the dynamics of valence parton could play an
major role in exciting a pion.

\section*{Acknowledgments}

We thank C.\ D.\ Roberts for helpful comments on the paper.
This material is based upon work supported by: (i) The U.S. Department
of Energy, Office of Science, Office of Nuclear Physics through the
Contract No. DE-SC0012704; (ii) The U.S.  Department of Energy,
Office of Science, Office of Nuclear Physics and Office of Advanced
Scientific Computing Research within the framework of Scientific
Discovery through Advance Computing (ScIDAC) award Computing the
Properties of Matter with Leadership Computing Resources;(iii) X.G.
is partially supported by the NSFC Grant Number 11890712.  (iv)
N.K. is supported by Jefferson Science Associates, LLC under U.S.
DOE Contract No. DE-AC05-06OR23177 and in part by U.S. DOE grant
No. DE-FG02-04ER41302.  (v) S.S. is supported by the National Science
Foundation under CAREER Award PHY-1847893 and by the RHIC Physics
Fellow Program of the RIKEN BNL Research Center (vi)  Y.Z. is
partially supported by the U.S. Department of Energy, Office of
Science, Office of Nuclear Physics, within the framework of the TMD
Topical Collaboration.  (vii) This research used awards of computer
time provided by the INCITE and ALCC programs at Oak Ridge Leadership
Computing Facility, a DOE Office of Science User Facility operated
under Contract No.  DE-AC05- 00OR22725. (viii) Computations for
this work were carried out in part on facilities of the USQCD
Collaboration, which are funded by the Office of Science of the
U.S. Department of Energy.

\bibliography{paper.bib}

\begin{thebibliography}{67}%
\makeatletter
\providecommand \@ifxundefined [1]{%
 \@ifx{#1\undefined}
}%
\providecommand \@ifnum [1]{%
 \ifnum #1\expandafter \@firstoftwo
 \else \expandafter \@secondoftwo
 \fi
}%
\providecommand \@ifx [1]{%
 \ifx #1\expandafter \@firstoftwo
 \else \expandafter \@secondoftwo
 \fi
}%
\providecommand \natexlab [1]{#1}%
\providecommand \enquote  [1]{``#1''}%
\providecommand \bibnamefont  [1]{#1}%
\providecommand \bibfnamefont [1]{#1}%
\providecommand \citenamefont [1]{#1}%
\providecommand \href@noop [0]{\@secondoftwo}%
\providecommand \href [0]{\begingroup \@sanitize@url \@href}%
\providecommand \@href[1]{\@@startlink{#1}\@@href}%
\providecommand \@@href[1]{\endgroup#1\@@endlink}%
\providecommand \@sanitize@url [0]{\catcode `\\12\catcode `\$12\catcode
  `\&12\catcode `\#12\catcode `\^12\catcode `\_12\catcode `\%12\relax}%
\providecommand \@@startlink[1]{}%
\providecommand \@@endlink[0]{}%
\providecommand \url  [0]{\begingroup\@sanitize@url \@url }%
\providecommand \@url [1]{\endgroup\@href {#1}{\urlprefix }}%
\providecommand \urlprefix  [0]{URL }%
\providecommand \Eprint [0]{\href }%
\providecommand \doibase [0]{http://dx.doi.org/}%
\providecommand \selectlanguage [0]{\@gobble}%
\providecommand \bibinfo  [0]{\@secondoftwo}%
\providecommand \bibfield  [0]{\@secondoftwo}%
\providecommand \translation [1]{[#1]}%
\providecommand \BibitemOpen [0]{}%
\providecommand \bibitemStop [0]{}%
\providecommand \bibitemNoStop [0]{.\EOS\space}%
\providecommand \EOS [0]{\spacefactor3000\relax}%
\providecommand \BibitemShut  [1]{\csname bibitem#1\endcsname}%
\let\auto@bib@innerbib\@empty
%</preamble>
\bibitem [{\citenamefont {Badier}\ \emph {et~al.}(1983)\citenamefont {Badier}
  \emph {et~al.}}]{Badier:1983mj}%
  \BibitemOpen
  \bibfield  {author} {\bibinfo {author} {\bibfnamefont {J.}~\bibnamefont
  {Badier}} \emph {et~al.} (\bibinfo {collaboration} {NA3}),\ }\href {\doibase
  10.1007/BF01573728} {\bibfield  {journal} {\bibinfo  {journal} {Z. Phys. C}\
  }\textbf {\bibinfo {volume} {18}},\ \bibinfo {pages} {281} (\bibinfo {year}
  {1983})}\BibitemShut {NoStop}%
\bibitem [{\citenamefont {Betev}\ \emph {et~al.}(1985)\citenamefont {Betev}
  \emph {et~al.}}]{Betev:1985pf}%
  \BibitemOpen
  \bibfield  {author} {\bibinfo {author} {\bibfnamefont {B.}~\bibnamefont
  {Betev}} \emph {et~al.} (\bibinfo {collaboration} {NA10}),\ }\href {\doibase
  10.1007/BF01550243} {\bibfield  {journal} {\bibinfo  {journal} {Z. Phys. C}\
  }\textbf {\bibinfo {volume} {28}},\ \bibinfo {pages} {9} (\bibinfo {year}
  {1985})}\BibitemShut {NoStop}%
\bibitem [{\citenamefont {Conway}\ \emph {et~al.}(1989)\citenamefont {Conway}
  \emph {et~al.}}]{Conway:1989fs}%
  \BibitemOpen
  \bibfield  {author} {\bibinfo {author} {\bibfnamefont {J.}~\bibnamefont
  {Conway}} \emph {et~al.},\ }\href {\doibase 10.1103/PhysRevD.39.92}
  {\bibfield  {journal} {\bibinfo  {journal} {Phys. Rev. D}\ }\textbf {\bibinfo
  {volume} {39}},\ \bibinfo {pages} {92} (\bibinfo {year} {1989})}\BibitemShut
  {NoStop}%
\bibitem [{\citenamefont {Owens}(1984)}]{Owens:1984zj}%
  \BibitemOpen
  \bibfield  {author} {\bibinfo {author} {\bibfnamefont {J.}~\bibnamefont
  {Owens}},\ }\href {\doibase 10.1103/PhysRevD.30.943} {\bibfield  {journal}
  {\bibinfo  {journal} {Phys. Rev. D}\ }\textbf {\bibinfo {volume} {30}},\
  \bibinfo {pages} {943} (\bibinfo {year} {1984})}\BibitemShut {NoStop}%
\bibitem [{\citenamefont {Sutton}\ \emph {et~al.}(1992)\citenamefont {Sutton},
  \citenamefont {Martin}, \citenamefont {Roberts},\ and\ \citenamefont
  {Stirling}}]{Sutton:1991ay}%
  \BibitemOpen
  \bibfield  {author} {\bibinfo {author} {\bibfnamefont {P.}~\bibnamefont
  {Sutton}}, \bibinfo {author} {\bibfnamefont {A.~D.}\ \bibnamefont {Martin}},
  \bibinfo {author} {\bibfnamefont {R.}~\bibnamefont {Roberts}}, \ and\
  \bibinfo {author} {\bibfnamefont {W.}~\bibnamefont {Stirling}},\ }\href
  {\doibase 10.1103/PhysRevD.45.2349} {\bibfield  {journal} {\bibinfo
  {journal} {Phys. Rev. D}\ }\textbf {\bibinfo {volume} {45}},\ \bibinfo
  {pages} {2349} (\bibinfo {year} {1992})}\BibitemShut {NoStop}%
\bibitem [{\citenamefont {Gluck}\ \emph {et~al.}(1992)\citenamefont {Gluck},
  \citenamefont {Reya},\ and\ \citenamefont {Vogt}}]{Gluck:1991ey}%
  \BibitemOpen
  \bibfield  {author} {\bibinfo {author} {\bibfnamefont {M.}~\bibnamefont
  {Gluck}}, \bibinfo {author} {\bibfnamefont {E.}~\bibnamefont {Reya}}, \ and\
  \bibinfo {author} {\bibfnamefont {A.}~\bibnamefont {Vogt}},\ }\href {\doibase
  10.1007/BF01559743} {\bibfield  {journal} {\bibinfo  {journal} {Z. Phys. C}\
  }\textbf {\bibinfo {volume} {53}},\ \bibinfo {pages} {651} (\bibinfo {year}
  {1992})}\BibitemShut {NoStop}%
\bibitem [{\citenamefont {Gluck}\ \emph {et~al.}(1999)\citenamefont {Gluck},
  \citenamefont {Reya},\ and\ \citenamefont {Schienbein}}]{Gluck:1999xe}%
  \BibitemOpen
  \bibfield  {author} {\bibinfo {author} {\bibfnamefont {M.}~\bibnamefont
  {Gluck}}, \bibinfo {author} {\bibfnamefont {E.}~\bibnamefont {Reya}}, \ and\
  \bibinfo {author} {\bibfnamefont {I.}~\bibnamefont {Schienbein}},\ }\href
  {\doibase 10.1007/s100529900124} {\bibfield  {journal} {\bibinfo  {journal}
  {Eur. Phys. J. C}\ }\textbf {\bibinfo {volume} {10}},\ \bibinfo {pages} {313}
  (\bibinfo {year} {1999})},\ \Eprint {http://arxiv.org/abs/hep-ph/9903288}
  {arXiv:hep-ph/9903288} \BibitemShut {NoStop}%
\bibitem [{\citenamefont {Wijesooriya}\ \emph {et~al.}(2005)\citenamefont
  {Wijesooriya}, \citenamefont {Reimer},\ and\ \citenamefont
  {Holt}}]{Wijesooriya:2005ir}%
  \BibitemOpen
  \bibfield  {author} {\bibinfo {author} {\bibfnamefont {K.}~\bibnamefont
  {Wijesooriya}}, \bibinfo {author} {\bibfnamefont {P.}~\bibnamefont {Reimer}},
  \ and\ \bibinfo {author} {\bibfnamefont {R.}~\bibnamefont {Holt}},\ }\href
  {\doibase 10.1103/PhysRevC.72.065203} {\bibfield  {journal} {\bibinfo
  {journal} {Phys. Rev. C}\ }\textbf {\bibinfo {volume} {72}},\ \bibinfo
  {pages} {065203} (\bibinfo {year} {2005})},\ \Eprint
  {http://arxiv.org/abs/nucl-ex/0509012} {arXiv:nucl-ex/0509012} \BibitemShut
  {NoStop}%
\bibitem [{\citenamefont {Barry}\ \emph {et~al.}(2018)\citenamefont {Barry},
  \citenamefont {Sato}, \citenamefont {Melnitchouk},\ and\ \citenamefont
  {Ji}}]{Barry:2018ort}%
  \BibitemOpen
  \bibfield  {author} {\bibinfo {author} {\bibfnamefont {P.}~\bibnamefont
  {Barry}}, \bibinfo {author} {\bibfnamefont {N.}~\bibnamefont {Sato}},
  \bibinfo {author} {\bibfnamefont {W.}~\bibnamefont {Melnitchouk}}, \ and\
  \bibinfo {author} {\bibfnamefont {C.-R.}\ \bibnamefont {Ji}},\ }\href
  {\doibase 10.1103/PhysRevLett.121.152001} {\bibfield  {journal} {\bibinfo
  {journal} {Phys. Rev. Lett.}\ }\textbf {\bibinfo {volume} {121}},\ \bibinfo
  {pages} {152001} (\bibinfo {year} {2018})},\ \Eprint
  {http://arxiv.org/abs/1804.01965} {arXiv:1804.01965 [hep-ph]} \BibitemShut
  {NoStop}%
\bibitem [{\citenamefont {Novikov}\ \emph {et~al.}(2020)\citenamefont {Novikov}
  \emph {et~al.}}]{Novikov:2020snp}%
  \BibitemOpen
  \bibfield  {author} {\bibinfo {author} {\bibfnamefont {I.}~\bibnamefont
  {Novikov}} \emph {et~al.},\ }\href {\doibase 10.1103/PhysRevD.102.014040}
  {\bibfield  {journal} {\bibinfo  {journal} {Phys. Rev. D}\ }\textbf {\bibinfo
  {volume} {102}},\ \bibinfo {pages} {014040} (\bibinfo {year} {2020})},\
  \Eprint {http://arxiv.org/abs/2002.02902} {arXiv:2002.02902 [hep-ph]}
  \BibitemShut {NoStop}%
\bibitem [{\citenamefont {Aicher}\ \emph {et~al.}(2010)\citenamefont {Aicher},
  \citenamefont {Schafer},\ and\ \citenamefont {Vogelsang}}]{Aicher:2010cb}%
  \BibitemOpen
  \bibfield  {author} {\bibinfo {author} {\bibfnamefont {M.}~\bibnamefont
  {Aicher}}, \bibinfo {author} {\bibfnamefont {A.}~\bibnamefont {Schafer}}, \
  and\ \bibinfo {author} {\bibfnamefont {W.}~\bibnamefont {Vogelsang}},\ }\href
  {\doibase 10.1103/PhysRevLett.105.252003} {\bibfield  {journal} {\bibinfo
  {journal} {Phys. Rev. Lett.}\ }\textbf {\bibinfo {volume} {105}},\ \bibinfo
  {pages} {252003} (\bibinfo {year} {2010})},\ \Eprint
  {http://arxiv.org/abs/1009.2481} {arXiv:1009.2481 [hep-ph]} \BibitemShut
  {NoStop}%
\bibitem [{\citenamefont {Nguyen}\ \emph {et~al.}(2011)\citenamefont {Nguyen},
  \citenamefont {Bashir}, \citenamefont {Roberts},\ and\ \citenamefont
  {Tandy}}]{Nguyen:2011jy}%
  \BibitemOpen
  \bibfield  {author} {\bibinfo {author} {\bibfnamefont {T.}~\bibnamefont
  {Nguyen}}, \bibinfo {author} {\bibfnamefont {A.}~\bibnamefont {Bashir}},
  \bibinfo {author} {\bibfnamefont {C.~D.}\ \bibnamefont {Roberts}}, \ and\
  \bibinfo {author} {\bibfnamefont {P.~C.}\ \bibnamefont {Tandy}},\ }\href
  {\doibase 10.1103/PhysRevC.83.062201} {\bibfield  {journal} {\bibinfo
  {journal} {Phys. Rev. C}\ }\textbf {\bibinfo {volume} {83}},\ \bibinfo
  {pages} {062201} (\bibinfo {year} {2011})},\ \Eprint
  {http://arxiv.org/abs/1102.2448} {arXiv:1102.2448 [nucl-th]} \BibitemShut
  {NoStop}%
\bibitem [{\citenamefont {Chen}\ \emph
  {et~al.}(2016{\natexlab{a}})\citenamefont {Chen}, \citenamefont {Chang},
  \citenamefont {Roberts}, \citenamefont {Wan},\ and\ \citenamefont
  {Zong}}]{Chen:2016sno}%
  \BibitemOpen
  \bibfield  {author} {\bibinfo {author} {\bibfnamefont {C.}~\bibnamefont
  {Chen}}, \bibinfo {author} {\bibfnamefont {L.}~\bibnamefont {Chang}},
  \bibinfo {author} {\bibfnamefont {C.~D.}\ \bibnamefont {Roberts}}, \bibinfo
  {author} {\bibfnamefont {S.}~\bibnamefont {Wan}}, \ and\ \bibinfo {author}
  {\bibfnamefont {H.-S.}\ \bibnamefont {Zong}},\ }\href {\doibase
  10.1103/PhysRevD.93.074021} {\bibfield  {journal} {\bibinfo  {journal} {Phys.
  Rev. D}\ }\textbf {\bibinfo {volume} {93}},\ \bibinfo {pages} {074021}
  (\bibinfo {year} {2016}{\natexlab{a}})},\ \Eprint
  {http://arxiv.org/abs/1602.01502} {arXiv:1602.01502 [nucl-th]} \BibitemShut
  {NoStop}%
\bibitem [{\citenamefont {Cui}\ \emph {et~al.}(2020)\citenamefont {Cui},
  \citenamefont {Ding}, \citenamefont {Gao}, \citenamefont {Raya},
  \citenamefont {Binosi}, \citenamefont {Chang}, \citenamefont {Roberts},
  \citenamefont {Rodr\'\i{}guez-Quintero},\ and\ \citenamefont
  {Schmidt}}]{Cui:2020tdf}%
  \BibitemOpen
  \bibfield  {author} {\bibinfo {author} {\bibfnamefont {Z.-F.}\ \bibnamefont
  {Cui}}, \bibinfo {author} {\bibfnamefont {M.}~\bibnamefont {Ding}}, \bibinfo
  {author} {\bibfnamefont {F.}~\bibnamefont {Gao}}, \bibinfo {author}
  {\bibfnamefont {K.}~\bibnamefont {Raya}}, \bibinfo {author} {\bibfnamefont
  {D.}~\bibnamefont {Binosi}}, \bibinfo {author} {\bibfnamefont
  {L.}~\bibnamefont {Chang}}, \bibinfo {author} {\bibfnamefont {C.~D.}\
  \bibnamefont {Roberts}}, \bibinfo {author} {\bibfnamefont {J.}~\bibnamefont
  {Rodr\'\i{}guez-Quintero}}, \ and\ \bibinfo {author} {\bibfnamefont {S.~M.}\
  \bibnamefont {Schmidt}},\ }\href {\doibase 10.1140/epjc/s10052-020-08578-4}
  {\bibfield  {journal} {\bibinfo  {journal} {Eur. Phys. J. C}\ }\textbf
  {\bibinfo {volume} {80}},\ \bibinfo {pages} {1064} (\bibinfo {year}
  {2020})}\BibitemShut {NoStop}%
\bibitem [{\citenamefont {Roberts}\ and\ \citenamefont
  {Schmidt}(2020)}]{Roberts:2020udq}%
  \BibitemOpen
  \bibfield  {author} {\bibinfo {author} {\bibfnamefont {C.~D.}\ \bibnamefont
  {Roberts}}\ and\ \bibinfo {author} {\bibfnamefont {S.~M.}\ \bibnamefont
  {Schmidt}}\ }(\bibinfo {year} {2020})\ \Eprint
  {http://arxiv.org/abs/2006.08782} {arXiv:2006.08782 [hep-ph]} \BibitemShut
  {NoStop}%
\bibitem [{\citenamefont {de~Teramond}\ \emph {et~al.}(2018)\citenamefont
  {de~Teramond}, \citenamefont {Liu}, \citenamefont {Sufian}, \citenamefont
  {Dosch}, \citenamefont {Brodsky},\ and\ \citenamefont
  {Deur}}]{deTeramond:2018ecg}%
  \BibitemOpen
  \bibfield  {author} {\bibinfo {author} {\bibfnamefont {G.~F.}\ \bibnamefont
  {de~Teramond}}, \bibinfo {author} {\bibfnamefont {T.}~\bibnamefont {Liu}},
  \bibinfo {author} {\bibfnamefont {R.~S.}\ \bibnamefont {Sufian}}, \bibinfo
  {author} {\bibfnamefont {H.~G.}\ \bibnamefont {Dosch}}, \bibinfo {author}
  {\bibfnamefont {S.~J.}\ \bibnamefont {Brodsky}}, \ and\ \bibinfo {author}
  {\bibfnamefont {A.}~\bibnamefont {Deur}} (\bibinfo {collaboration} {HLFHS}),\
  }\href {\doibase 10.1103/PhysRevLett.120.182001} {\bibfield  {journal}
  {\bibinfo  {journal} {Phys. Rev. Lett.}\ }\textbf {\bibinfo {volume} {120}},\
  \bibinfo {pages} {182001} (\bibinfo {year} {2018})},\ \Eprint
  {http://arxiv.org/abs/1801.09154} {arXiv:1801.09154 [hep-ph]} \BibitemShut
  {NoStop}%
\bibitem [{\citenamefont {Ruiz~Arriola}(2002)}]{RuizArriola:2002wr}%
  \BibitemOpen
  \bibfield  {author} {\bibinfo {author} {\bibfnamefont {E.}~\bibnamefont
  {Ruiz~Arriola}},\ }\href@noop {} {\bibfield  {journal} {\bibinfo  {journal}
  {Acta Phys. Polon. B}\ }\textbf {\bibinfo {volume} {33}},\ \bibinfo {pages}
  {4443} (\bibinfo {year} {2002})},\ \Eprint
  {http://arxiv.org/abs/hep-ph/0210007} {arXiv:hep-ph/0210007} \BibitemShut
  {NoStop}%
\bibitem [{\citenamefont {Broniowski}\ and\ \citenamefont
  {Ruiz~Arriola}(2017)}]{Broniowski:2017wbr}%
  \BibitemOpen
  \bibfield  {author} {\bibinfo {author} {\bibfnamefont {W.}~\bibnamefont
  {Broniowski}}\ and\ \bibinfo {author} {\bibfnamefont {E.}~\bibnamefont
  {Ruiz~Arriola}},\ }\href {\doibase 10.1016/j.physletb.2017.08.055} {\bibfield
   {journal} {\bibinfo  {journal} {Phys. Lett. B}\ }\textbf {\bibinfo {volume}
  {773}},\ \bibinfo {pages} {385} (\bibinfo {year} {2017})},\ \Eprint
  {http://arxiv.org/abs/1707.09588} {arXiv:1707.09588 [hep-ph]} \BibitemShut
  {NoStop}%
\bibitem [{\citenamefont {Lan}\ \emph {et~al.}(2020)\citenamefont {Lan},
  \citenamefont {Mondal}, \citenamefont {Jia}, \citenamefont {Zhao},\ and\
  \citenamefont {Vary}}]{Lan:2019rba}%
  \BibitemOpen
  \bibfield  {author} {\bibinfo {author} {\bibfnamefont {J.}~\bibnamefont
  {Lan}}, \bibinfo {author} {\bibfnamefont {C.}~\bibnamefont {Mondal}},
  \bibinfo {author} {\bibfnamefont {S.}~\bibnamefont {Jia}}, \bibinfo {author}
  {\bibfnamefont {X.}~\bibnamefont {Zhao}}, \ and\ \bibinfo {author}
  {\bibfnamefont {J.~P.}\ \bibnamefont {Vary}},\ }\href {\doibase
  10.1103/PhysRevD.101.034024} {\bibfield  {journal} {\bibinfo  {journal}
  {Phys. Rev. D}\ }\textbf {\bibinfo {volume} {101}},\ \bibinfo {pages}
  {034024} (\bibinfo {year} {2020})},\ \Eprint
  {http://arxiv.org/abs/1907.01509} {arXiv:1907.01509 [nucl-th]} \BibitemShut
  {NoStop}%
\bibitem [{\citenamefont {Bednar}\ \emph {et~al.}(2020)\citenamefont {Bednar},
  \citenamefont {Clo\"et},\ and\ \citenamefont {Tandy}}]{Bednar:2018mtf}%
  \BibitemOpen
  \bibfield  {author} {\bibinfo {author} {\bibfnamefont {K.~D.}\ \bibnamefont
  {Bednar}}, \bibinfo {author} {\bibfnamefont {I.~C.}\ \bibnamefont {Clo\"et}},
  \ and\ \bibinfo {author} {\bibfnamefont {P.~C.}\ \bibnamefont {Tandy}},\
  }\href {\doibase 10.1103/PhysRevLett.124.042002} {\bibfield  {journal}
  {\bibinfo  {journal} {Phys. Rev. Lett.}\ }\textbf {\bibinfo {volume} {124}},\
  \bibinfo {pages} {042002} (\bibinfo {year} {2020})},\ \Eprint
  {http://arxiv.org/abs/1811.12310} {arXiv:1811.12310 [nucl-th]} \BibitemShut
  {NoStop}%
\bibitem [{\citenamefont {Aguilar}\ \emph {et~al.}(2019)\citenamefont {Aguilar}
  \emph {et~al.}}]{Aguilar:2019teb}%
  \BibitemOpen
  \bibfield  {author} {\bibinfo {author} {\bibfnamefont {A.~C.}\ \bibnamefont
  {Aguilar}} \emph {et~al.},\ }\href {\doibase 10.1140/epja/i2019-12885-0}
  {\bibfield  {journal} {\bibinfo  {journal} {Eur. Phys. J. A}\ }\textbf
  {\bibinfo {volume} {55}},\ \bibinfo {pages} {190} (\bibinfo {year} {2019})},\
  \Eprint {http://arxiv.org/abs/1907.08218} {arXiv:1907.08218 [nucl-ex]}
  \BibitemShut {NoStop}%
\bibitem [{\citenamefont {Adams}\ \emph {et~al.}(2018)\citenamefont {Adams}
  \emph {et~al.}}]{Denisov:2018unj}%
  \BibitemOpen
  \bibfield  {author} {\bibinfo {author} {\bibfnamefont {B.}~\bibnamefont
  {Adams}} \emph {et~al.},\ }\href@noop {} {\  (\bibinfo {year} {2018})},\
  \Eprint {http://arxiv.org/abs/1808.00848} {arXiv:1808.00848 [hep-ex]}
  \BibitemShut {NoStop}%
\bibitem [{\citenamefont {Ji}(2013)}]{Ji:2013dva}%
  \BibitemOpen
  \bibfield  {author} {\bibinfo {author} {\bibfnamefont {X.}~\bibnamefont
  {Ji}},\ }\href {\doibase 10.1103/PhysRevLett.110.262002} {\bibfield
  {journal} {\bibinfo  {journal} {Phys. Rev. Lett.}\ }\textbf {\bibinfo
  {volume} {110}},\ \bibinfo {pages} {262002} (\bibinfo {year} {2013})},\
  \Eprint {http://arxiv.org/abs/1305.1539} {arXiv:1305.1539 [hep-ph]}
  \BibitemShut {NoStop}%
\bibitem [{\citenamefont {Ji}(2014)}]{Ji:2014gla}%
  \BibitemOpen
  \bibfield  {author} {\bibinfo {author} {\bibfnamefont {X.}~\bibnamefont
  {Ji}},\ }\href {\doibase 10.1007/s11433-014-5492-3} {\bibfield  {journal}
  {\bibinfo  {journal} {Sci. China Phys. Mech. Astron.}\ }\textbf {\bibinfo
  {volume} {57}},\ \bibinfo {pages} {1407} (\bibinfo {year} {2014})},\ \Eprint
  {http://arxiv.org/abs/1404.6680} {arXiv:1404.6680 [hep-ph]} \BibitemShut
  {NoStop}%
\bibitem [{\citenamefont
  {Radyushkin}(2017{\natexlab{a}})}]{Radyushkin:2017cyf}%
  \BibitemOpen
  \bibfield  {author} {\bibinfo {author} {\bibfnamefont {A.}~\bibnamefont
  {Radyushkin}},\ }\href {\doibase 10.1103/PhysRevD.96.034025} {\bibfield
  {journal} {\bibinfo  {journal} {Phys. Rev. D}\ }\textbf {\bibinfo {volume}
  {96}},\ \bibinfo {pages} {034025} (\bibinfo {year} {2017}{\natexlab{a}})},\
  \Eprint {http://arxiv.org/abs/1705.01488} {arXiv:1705.01488 [hep-ph]}
  \BibitemShut {NoStop}%
\bibitem [{\citenamefont {Orginos}\ \emph {et~al.}(2017)\citenamefont
  {Orginos}, \citenamefont {Radyushkin}, \citenamefont {Karpie},\ and\
  \citenamefont {Zafeiropoulos}}]{Orginos:2017kos}%
  \BibitemOpen
  \bibfield  {author} {\bibinfo {author} {\bibfnamefont {K.}~\bibnamefont
  {Orginos}}, \bibinfo {author} {\bibfnamefont {A.}~\bibnamefont {Radyushkin}},
  \bibinfo {author} {\bibfnamefont {J.}~\bibnamefont {Karpie}}, \ and\ \bibinfo
  {author} {\bibfnamefont {S.}~\bibnamefont {Zafeiropoulos}},\ }\href {\doibase
  10.1103/PhysRevD.96.094503} {\bibfield  {journal} {\bibinfo  {journal} {Phys.
  Rev. D}\ }\textbf {\bibinfo {volume} {96}},\ \bibinfo {pages} {094503}
  (\bibinfo {year} {2017})},\ \Eprint {http://arxiv.org/abs/1706.05373}
  {arXiv:1706.05373 [hep-ph]} \BibitemShut {NoStop}%
\bibitem [{\citenamefont {Braun}\ and\ \citenamefont
  {M\"uller}(2008)}]{Braun:2007wv}%
  \BibitemOpen
  \bibfield  {author} {\bibinfo {author} {\bibfnamefont {V.}~\bibnamefont
  {Braun}}\ and\ \bibinfo {author} {\bibfnamefont {D.}~\bibnamefont
  {M\"uller}},\ }\href {\doibase 10.1140/epjc/s10052-008-0608-4} {\bibfield
  {journal} {\bibinfo  {journal} {Eur. Phys. J. C}\ }\textbf {\bibinfo {volume}
  {55}},\ \bibinfo {pages} {349} (\bibinfo {year} {2008})},\ \Eprint
  {http://arxiv.org/abs/0709.1348} {arXiv:0709.1348 [hep-ph]} \BibitemShut
  {NoStop}%
\bibitem [{\citenamefont {Ma}\ and\ \citenamefont
  {Qiu}(2018{\natexlab{a}})}]{Ma:2014jla}%
  \BibitemOpen
  \bibfield  {author} {\bibinfo {author} {\bibfnamefont {Y.-Q.}\ \bibnamefont
  {Ma}}\ and\ \bibinfo {author} {\bibfnamefont {J.-W.}\ \bibnamefont {Qiu}},\
  }\href {\doibase 10.1103/PhysRevD.98.074021} {\bibfield  {journal} {\bibinfo
  {journal} {Phys. Rev. D}\ }\textbf {\bibinfo {volume} {98}},\ \bibinfo
  {pages} {074021} (\bibinfo {year} {2018}{\natexlab{a}})},\ \Eprint
  {http://arxiv.org/abs/1404.6860} {arXiv:1404.6860 [hep-ph]} \BibitemShut
  {NoStop}%
\bibitem [{\citenamefont {Ma}\ and\ \citenamefont
  {Qiu}(2018{\natexlab{b}})}]{Ma:2017pxb}%
  \BibitemOpen
  \bibfield  {author} {\bibinfo {author} {\bibfnamefont {Y.-Q.}\ \bibnamefont
  {Ma}}\ and\ \bibinfo {author} {\bibfnamefont {J.-W.}\ \bibnamefont {Qiu}},\
  }\href {\doibase 10.1103/PhysRevLett.120.022003} {\bibfield  {journal}
  {\bibinfo  {journal} {Phys. Rev. Lett.}\ }\textbf {\bibinfo {volume} {120}},\
  \bibinfo {pages} {022003} (\bibinfo {year} {2018}{\natexlab{b}})},\ \Eprint
  {http://arxiv.org/abs/1709.03018} {arXiv:1709.03018 [hep-ph]} \BibitemShut
  {NoStop}%
\bibitem [{\citenamefont {Constantinou}(2020)}]{Constantinou:2020pek}%
  \BibitemOpen
  \bibfield  {author} {\bibinfo {author} {\bibfnamefont {M.}~\bibnamefont
  {Constantinou}},\ }in\ \href@noop {} {\emph {\bibinfo {booktitle} {{38th
  International Symposium on Lattice Field Theory}}}}\ (\bibinfo {year}
  {2020})\ \Eprint {http://arxiv.org/abs/2010.02445} {arXiv:2010.02445
  [hep-lat]} \BibitemShut {NoStop}%
\bibitem [{\citenamefont {Zhao}(2019)}]{Zhao:2018fyu}%
  \BibitemOpen
  \bibfield  {author} {\bibinfo {author} {\bibfnamefont {Y.}~\bibnamefont
  {Zhao}},\ }\href {\doibase 10.1142/S0217751X18300338} {\bibfield  {journal}
  {\bibinfo  {journal} {Int. J. Mod. Phys. A}\ }\textbf {\bibinfo {volume}
  {33}},\ \bibinfo {pages} {1830033} (\bibinfo {year} {2019})},\ \Eprint
  {http://arxiv.org/abs/1812.07192} {arXiv:1812.07192 [hep-ph]} \BibitemShut
  {NoStop}%
\bibitem [{\citenamefont {Cichy}\ and\ \citenamefont
  {Constantinou}(2019)}]{Cichy:2018mum}%
  \BibitemOpen
  \bibfield  {author} {\bibinfo {author} {\bibfnamefont {K.}~\bibnamefont
  {Cichy}}\ and\ \bibinfo {author} {\bibfnamefont {M.}~\bibnamefont
  {Constantinou}},\ }\href {\doibase 10.1155/2019/3036904} {\bibfield
  {journal} {\bibinfo  {journal} {Adv. High Energy Phys.}\ }\textbf {\bibinfo
  {volume} {2019}},\ \bibinfo {pages} {3036904} (\bibinfo {year} {2019})},\
  \Eprint {http://arxiv.org/abs/1811.07248} {arXiv:1811.07248 [hep-lat]}
  \BibitemShut {NoStop}%
\bibitem [{\citenamefont {Monahan}(2018)}]{Monahan:2018euv}%
  \BibitemOpen
  \bibfield  {author} {\bibinfo {author} {\bibfnamefont {C.}~\bibnamefont
  {Monahan}},\ }\href {\doibase 10.22323/1.334.0018} {\bibfield  {journal}
  {\bibinfo  {journal} {PoS}\ }\textbf {\bibinfo {volume} {LATTICE2018}},\
  \bibinfo {pages} {018} (\bibinfo {year} {2018})},\ \Eprint
  {http://arxiv.org/abs/1811.00678} {arXiv:1811.00678 [hep-lat]} \BibitemShut
  {NoStop}%
\bibitem [{\citenamefont {Ji}\ \emph {et~al.}(2020{\natexlab{a}})\citenamefont
  {Ji}, \citenamefont {Liu}, \citenamefont {Liu}, \citenamefont {Zhang},\ and\
  \citenamefont {Zhao}}]{Ji:2020ect}%
  \BibitemOpen
  \bibfield  {author} {\bibinfo {author} {\bibfnamefont {X.}~\bibnamefont
  {Ji}}, \bibinfo {author} {\bibfnamefont {Y.-S.}\ \bibnamefont {Liu}},
  \bibinfo {author} {\bibfnamefont {Y.}~\bibnamefont {Liu}}, \bibinfo {author}
  {\bibfnamefont {J.-H.}\ \bibnamefont {Zhang}}, \ and\ \bibinfo {author}
  {\bibfnamefont {Y.}~\bibnamefont {Zhao}},\ }\href@noop {} {\  (\bibinfo
  {year} {2020}{\natexlab{a}})},\ \Eprint {http://arxiv.org/abs/2004.03543}
  {arXiv:2004.03543 [hep-ph]} \BibitemShut {NoStop}%
\bibitem [{\citenamefont {Gao}\ \emph {et~al.}(2020)\citenamefont {Gao},
  \citenamefont {Jin}, \citenamefont {Kallidonis}, \citenamefont {Karthik},
  \citenamefont {Mukherjee}, \citenamefont {Petreczky}, \citenamefont
  {Shugert}, \citenamefont {Syritsyn},\ and\ \citenamefont
  {Zhao}}]{Gao:2020ito}%
  \BibitemOpen
  \bibfield  {author} {\bibinfo {author} {\bibfnamefont {X.}~\bibnamefont
  {Gao}}, \bibinfo {author} {\bibfnamefont {L.}~\bibnamefont {Jin}}, \bibinfo
  {author} {\bibfnamefont {C.}~\bibnamefont {Kallidonis}}, \bibinfo {author}
  {\bibfnamefont {N.}~\bibnamefont {Karthik}}, \bibinfo {author} {\bibfnamefont
  {S.}~\bibnamefont {Mukherjee}}, \bibinfo {author} {\bibfnamefont
  {P.}~\bibnamefont {Petreczky}}, \bibinfo {author} {\bibfnamefont
  {C.}~\bibnamefont {Shugert}}, \bibinfo {author} {\bibfnamefont
  {S.}~\bibnamefont {Syritsyn}}, \ and\ \bibinfo {author} {\bibfnamefont
  {Y.}~\bibnamefont {Zhao}},\ }\href {\doibase 10.1103/PhysRevD.102.094513}
  {\bibfield  {journal} {\bibinfo  {journal} {Phys. Rev. D}\ }\textbf {\bibinfo
  {volume} {102}},\ \bibinfo {pages} {094513} (\bibinfo {year} {2020})},\
  \Eprint {http://arxiv.org/abs/2007.06590} {arXiv:2007.06590 [hep-lat]}
  \BibitemShut {NoStop}%
\bibitem [{\citenamefont {Zhang}\ \emph {et~al.}(2019)\citenamefont {Zhang},
  \citenamefont {Chen}, \citenamefont {Jin}, \citenamefont {Lin}, \citenamefont
  {Sch\"afer},\ and\ \citenamefont {Zhao}}]{Chen:2018fwa}%
  \BibitemOpen
  \bibfield  {author} {\bibinfo {author} {\bibfnamefont {J.-H.}\ \bibnamefont
  {Zhang}}, \bibinfo {author} {\bibfnamefont {J.-W.}\ \bibnamefont {Chen}},
  \bibinfo {author} {\bibfnamefont {L.}~\bibnamefont {Jin}}, \bibinfo {author}
  {\bibfnamefont {H.-W.}\ \bibnamefont {Lin}}, \bibinfo {author} {\bibfnamefont
  {A.}~\bibnamefont {Sch\"afer}}, \ and\ \bibinfo {author} {\bibfnamefont
  {Y.}~\bibnamefont {Zhao}},\ }\href {\doibase 10.1103/PhysRevD.100.034505}
  {\bibfield  {journal} {\bibinfo  {journal} {Phys. Rev. D}\ }\textbf {\bibinfo
  {volume} {100}},\ \bibinfo {pages} {034505} (\bibinfo {year} {2019})},\
  \Eprint {http://arxiv.org/abs/1804.01483} {arXiv:1804.01483 [hep-lat]}
  \BibitemShut {NoStop}%
\bibitem [{\citenamefont {Izubuchi}\ \emph {et~al.}(2019)\citenamefont
  {Izubuchi}, \citenamefont {Jin}, \citenamefont {Kallidonis}, \citenamefont
  {Karthik}, \citenamefont {Mukherjee}, \citenamefont {Petreczky},
  \citenamefont {Shugert},\ and\ \citenamefont {Syritsyn}}]{Izubuchi:2019lyk}%
  \BibitemOpen
  \bibfield  {author} {\bibinfo {author} {\bibfnamefont {T.}~\bibnamefont
  {Izubuchi}}, \bibinfo {author} {\bibfnamefont {L.}~\bibnamefont {Jin}},
  \bibinfo {author} {\bibfnamefont {C.}~\bibnamefont {Kallidonis}}, \bibinfo
  {author} {\bibfnamefont {N.}~\bibnamefont {Karthik}}, \bibinfo {author}
  {\bibfnamefont {S.}~\bibnamefont {Mukherjee}}, \bibinfo {author}
  {\bibfnamefont {P.}~\bibnamefont {Petreczky}}, \bibinfo {author}
  {\bibfnamefont {C.}~\bibnamefont {Shugert}}, \ and\ \bibinfo {author}
  {\bibfnamefont {S.}~\bibnamefont {Syritsyn}},\ }\href {\doibase
  10.1103/PhysRevD.100.034516} {\bibfield  {journal} {\bibinfo  {journal}
  {Phys. Rev. D}\ }\textbf {\bibinfo {volume} {100}},\ \bibinfo {pages}
  {034516} (\bibinfo {year} {2019})},\ \Eprint
  {http://arxiv.org/abs/1905.06349} {arXiv:1905.06349 [hep-lat]} \BibitemShut
  {NoStop}%
\bibitem [{\citenamefont {Jo\'o}\ \emph {et~al.}(2019)\citenamefont {Jo\'o},
  \citenamefont {Karpie}, \citenamefont {Orginos}, \citenamefont {Radyushkin},
  \citenamefont {Richards}, \citenamefont {Sufian},\ and\ \citenamefont
  {Zafeiropoulos}}]{Joo:2019bzr}%
  \BibitemOpen
  \bibfield  {author} {\bibinfo {author} {\bibfnamefont {B.}~\bibnamefont
  {Jo\'o}}, \bibinfo {author} {\bibfnamefont {J.}~\bibnamefont {Karpie}},
  \bibinfo {author} {\bibfnamefont {K.}~\bibnamefont {Orginos}}, \bibinfo
  {author} {\bibfnamefont {A.~V.}\ \bibnamefont {Radyushkin}}, \bibinfo
  {author} {\bibfnamefont {D.~G.}\ \bibnamefont {Richards}}, \bibinfo {author}
  {\bibfnamefont {R.~S.}\ \bibnamefont {Sufian}}, \ and\ \bibinfo {author}
  {\bibfnamefont {S.}~\bibnamefont {Zafeiropoulos}},\ }\href {\doibase
  10.1103/PhysRevD.100.114512} {\bibfield  {journal} {\bibinfo  {journal}
  {Phys. Rev. D}\ }\textbf {\bibinfo {volume} {100}},\ \bibinfo {pages}
  {114512} (\bibinfo {year} {2019})},\ \Eprint
  {http://arxiv.org/abs/1909.08517} {arXiv:1909.08517 [hep-lat]} \BibitemShut
  {NoStop}%
\bibitem [{\citenamefont {Lin}\ \emph {et~al.}(2020)\citenamefont {Lin},
  \citenamefont {Chen}, \citenamefont {Fan}, \citenamefont {Zhang},\ and\
  \citenamefont {Zhang}}]{Lin:2020ssv}%
  \BibitemOpen
  \bibfield  {author} {\bibinfo {author} {\bibfnamefont {H.-W.}\ \bibnamefont
  {Lin}}, \bibinfo {author} {\bibfnamefont {J.-W.}\ \bibnamefont {Chen}},
  \bibinfo {author} {\bibfnamefont {Z.}~\bibnamefont {Fan}}, \bibinfo {author}
  {\bibfnamefont {J.-H.}\ \bibnamefont {Zhang}}, \ and\ \bibinfo {author}
  {\bibfnamefont {R.}~\bibnamefont {Zhang}},\ }\href@noop {} {\  (\bibinfo
  {year} {2020})},\ \Eprint {http://arxiv.org/abs/2003.14128} {arXiv:2003.14128
  [hep-lat]} \BibitemShut {NoStop}%
\bibitem [{\citenamefont {Sufian}\ \emph {et~al.}(2019)\citenamefont {Sufian},
  \citenamefont {Karpie}, \citenamefont {Egerer}, \citenamefont {Orginos},
  \citenamefont {Qiu},\ and\ \citenamefont {Richards}}]{Sufian:2019bol}%
  \BibitemOpen
  \bibfield  {author} {\bibinfo {author} {\bibfnamefont {R.~S.}\ \bibnamefont
  {Sufian}}, \bibinfo {author} {\bibfnamefont {J.}~\bibnamefont {Karpie}},
  \bibinfo {author} {\bibfnamefont {C.}~\bibnamefont {Egerer}}, \bibinfo
  {author} {\bibfnamefont {K.}~\bibnamefont {Orginos}}, \bibinfo {author}
  {\bibfnamefont {J.-W.}\ \bibnamefont {Qiu}}, \ and\ \bibinfo {author}
  {\bibfnamefont {D.~G.}\ \bibnamefont {Richards}},\ }\href {\doibase
  10.1103/PhysRevD.99.074507} {\bibfield  {journal} {\bibinfo  {journal} {Phys.
  Rev. D}\ }\textbf {\bibinfo {volume} {99}},\ \bibinfo {pages} {074507}
  (\bibinfo {year} {2019})},\ \Eprint {http://arxiv.org/abs/1901.03921}
  {arXiv:1901.03921 [hep-lat]} \BibitemShut {NoStop}%
\bibitem [{\citenamefont {Sufian}\ \emph {et~al.}(2020)\citenamefont {Sufian},
  \citenamefont {Egerer}, \citenamefont {Karpie}, \citenamefont {Edwards},
  \citenamefont {Jo\'o}, \citenamefont {Ma}, \citenamefont {Orginos},
  \citenamefont {Qiu},\ and\ \citenamefont {Richards}}]{Sufian:2020vzb}%
  \BibitemOpen
  \bibfield  {author} {\bibinfo {author} {\bibfnamefont {R.~S.}\ \bibnamefont
  {Sufian}}, \bibinfo {author} {\bibfnamefont {C.}~\bibnamefont {Egerer}},
  \bibinfo {author} {\bibfnamefont {J.}~\bibnamefont {Karpie}}, \bibinfo
  {author} {\bibfnamefont {R.~G.}\ \bibnamefont {Edwards}}, \bibinfo {author}
  {\bibfnamefont {B.}~\bibnamefont {Jo\'o}}, \bibinfo {author} {\bibfnamefont
  {Y.-Q.}\ \bibnamefont {Ma}}, \bibinfo {author} {\bibfnamefont
  {K.}~\bibnamefont {Orginos}}, \bibinfo {author} {\bibfnamefont {J.-W.}\
  \bibnamefont {Qiu}}, \ and\ \bibinfo {author} {\bibfnamefont {D.~G.}\
  \bibnamefont {Richards}},\ }\href {\doibase 10.1103/PhysRevD.102.054508}
  {\bibfield  {journal} {\bibinfo  {journal} {Phys. Rev. D}\ }\textbf {\bibinfo
  {volume} {102}},\ \bibinfo {pages} {054508} (\bibinfo {year} {2020})},\
  \Eprint {http://arxiv.org/abs/2001.04960} {arXiv:2001.04960 [hep-lat]}
  \BibitemShut {NoStop}%
\bibitem [{\citenamefont {Karthik}(2021)}]{Karthik:2021qwz}%
  \BibitemOpen
  \bibfield  {author} {\bibinfo {author} {\bibfnamefont {N.}~\bibnamefont
  {Karthik}},\ }\href@noop {} {\  (\bibinfo {year} {2021})},\ \Eprint
  {http://arxiv.org/abs/2101.02224} {arXiv:2101.02224 [hep-lat]} \BibitemShut
  {NoStop}%
\bibitem [{\citenamefont {Braun}\ \emph {et~al.}(2019)\citenamefont {Braun},
  \citenamefont {Vladimirov},\ and\ \citenamefont {Zhang}}]{Braun:2018brg}%
  \BibitemOpen
  \bibfield  {author} {\bibinfo {author} {\bibfnamefont {V.~M.}\ \bibnamefont
  {Braun}}, \bibinfo {author} {\bibfnamefont {A.}~\bibnamefont {Vladimirov}}, \
  and\ \bibinfo {author} {\bibfnamefont {J.-H.}\ \bibnamefont {Zhang}},\ }\href
  {\doibase 10.1103/PhysRevD.99.014013} {\bibfield  {journal} {\bibinfo
  {journal} {Phys. Rev. D}\ }\textbf {\bibinfo {volume} {99}},\ \bibinfo
  {pages} {014013} (\bibinfo {year} {2019})},\ \Eprint
  {http://arxiv.org/abs/1810.00048} {arXiv:1810.00048 [hep-ph]} \BibitemShut
  {NoStop}%
\bibitem [{\citenamefont {Liu}\ and\ \citenamefont {Chen}(2020)}]{Liu:2020rqi}%
  \BibitemOpen
  \bibfield  {author} {\bibinfo {author} {\bibfnamefont {W.-Y.}\ \bibnamefont
  {Liu}}\ and\ \bibinfo {author} {\bibfnamefont {J.-W.}\ \bibnamefont {Chen}},\
  }\href@noop {} {\  (\bibinfo {year} {2020})},\ \Eprint
  {http://arxiv.org/abs/2010.06623} {arXiv:2010.06623 [hep-ph]} \BibitemShut
  {NoStop}%
\bibitem [{\citenamefont {Ji}\ \emph {et~al.}(2020{\natexlab{b}})\citenamefont
  {Ji}, \citenamefont {Liu}, \citenamefont {Sch\"afer}, \citenamefont {Wang},
  \citenamefont {Yang}, \citenamefont {Zhang},\ and\ \citenamefont
  {Zhao}}]{Ji:2020brr}%
  \BibitemOpen
  \bibfield  {author} {\bibinfo {author} {\bibfnamefont {X.}~\bibnamefont
  {Ji}}, \bibinfo {author} {\bibfnamefont {Y.}~\bibnamefont {Liu}}, \bibinfo
  {author} {\bibfnamefont {A.}~\bibnamefont {Sch\"afer}}, \bibinfo {author}
  {\bibfnamefont {W.}~\bibnamefont {Wang}}, \bibinfo {author} {\bibfnamefont
  {Y.-B.}\ \bibnamefont {Yang}}, \bibinfo {author} {\bibfnamefont {J.-H.}\
  \bibnamefont {Zhang}}, \ and\ \bibinfo {author} {\bibfnamefont
  {Y.}~\bibnamefont {Zhao}},\ }\href@noop {} {\  (\bibinfo {year}
  {2020}{\natexlab{b}})},\ \Eprint {http://arxiv.org/abs/2008.03886}
  {arXiv:2008.03886 [hep-ph]} \BibitemShut {NoStop}%
\bibitem [{\citenamefont {Roberts}(2020)}]{Roberts:2020hiw}%
  \BibitemOpen
  \bibfield  {author} {\bibinfo {author} {\bibfnamefont {C.~D.}\ \bibnamefont
  {Roberts}},\ }\href {\doibase 10.3390/sym12091468} {\bibfield  {journal}
  {\bibinfo  {journal} {Symmetry}\ }\textbf {\bibinfo {volume} {12}},\ \bibinfo
  {pages} {1468} (\bibinfo {year} {2020})},\ \Eprint
  {http://arxiv.org/abs/2009.04011} {arXiv:2009.04011 [hep-ph]} \BibitemShut
  {NoStop}%
\bibitem [{\citenamefont {McNeile}\ and\ \citenamefont
  {Michael}(2006)}]{McNeile:2006qy}%
  \BibitemOpen
  \bibfield  {author} {\bibinfo {author} {\bibfnamefont {C.}~\bibnamefont
  {McNeile}}\ and\ \bibinfo {author} {\bibfnamefont {C.}~\bibnamefont
  {Michael}} (\bibinfo {collaboration} {UKQCD}),\ }\href {\doibase
  10.1016/j.physletb.2006.09.056} {\bibfield  {journal} {\bibinfo  {journal}
  {Phys. Lett. B}\ }\textbf {\bibinfo {volume} {642}},\ \bibinfo {pages} {244}
  (\bibinfo {year} {2006})},\ \Eprint {http://arxiv.org/abs/hep-lat/0607032}
  {arXiv:hep-lat/0607032} \BibitemShut {NoStop}%
\bibitem [{\citenamefont {Mastropas}\ and\ \citenamefont
  {Richards}(2014)}]{Mastropas:2014fsa}%
  \BibitemOpen
  \bibfield  {author} {\bibinfo {author} {\bibfnamefont {E.~V.}\ \bibnamefont
  {Mastropas}}\ and\ \bibinfo {author} {\bibfnamefont {D.~G.}\ \bibnamefont
  {Richards}} (\bibinfo {collaboration} {Hadron Spectrum}),\ }\href {\doibase
  10.1103/PhysRevD.90.014511} {\bibfield  {journal} {\bibinfo  {journal} {Phys.
  Rev. D}\ }\textbf {\bibinfo {volume} {90}},\ \bibinfo {pages} {014511}
  (\bibinfo {year} {2014})},\ \Eprint {http://arxiv.org/abs/1403.5575}
  {arXiv:1403.5575 [hep-lat]} \BibitemShut {NoStop}%
\bibitem [{\citenamefont {Holl}\ \emph {et~al.}(2004)\citenamefont {Holl},
  \citenamefont {Krassnigg},\ and\ \citenamefont {Roberts}}]{Holl:2004fr}%
  \BibitemOpen
  \bibfield  {author} {\bibinfo {author} {\bibfnamefont {A.}~\bibnamefont
  {Holl}}, \bibinfo {author} {\bibfnamefont {A.}~\bibnamefont {Krassnigg}}, \
  and\ \bibinfo {author} {\bibfnamefont {C.~D.}\ \bibnamefont {Roberts}},\
  }\href {\doibase 10.1103/PhysRevC.70.042203} {\bibfield  {journal} {\bibinfo
  {journal} {Phys. Rev. C}\ }\textbf {\bibinfo {volume} {70}},\ \bibinfo
  {pages} {042203} (\bibinfo {year} {2004})},\ \Eprint
  {http://arxiv.org/abs/nucl-th/0406030} {arXiv:nucl-th/0406030} \BibitemShut
  {NoStop}%
\bibitem [{\citenamefont {Li}\ \emph {et~al.}(2016)\citenamefont {Li},
  \citenamefont {Chang}, \citenamefont {Gao}, \citenamefont {Roberts},
  \citenamefont {Schmidt},\ and\ \citenamefont {Zong}}]{Li:2016dzv}%
  \BibitemOpen
  \bibfield  {author} {\bibinfo {author} {\bibfnamefont {B.~L.}\ \bibnamefont
  {Li}}, \bibinfo {author} {\bibfnamefont {L.}~\bibnamefont {Chang}}, \bibinfo
  {author} {\bibfnamefont {F.}~\bibnamefont {Gao}}, \bibinfo {author}
  {\bibfnamefont {C.~D.}\ \bibnamefont {Roberts}}, \bibinfo {author}
  {\bibfnamefont {S.~M.}\ \bibnamefont {Schmidt}}, \ and\ \bibinfo {author}
  {\bibfnamefont {H.~S.}\ \bibnamefont {Zong}},\ }\href {\doibase
  10.1103/PhysRevD.93.114033} {\bibfield  {journal} {\bibinfo  {journal} {Phys.
  Rev. D}\ }\textbf {\bibinfo {volume} {93}},\ \bibinfo {pages} {114033}
  (\bibinfo {year} {2016})},\ \Eprint {http://arxiv.org/abs/1604.07415}
  {arXiv:1604.07415 [nucl-th]} \BibitemShut {NoStop}%
\bibitem [{\citenamefont {Chai}\ \emph {et~al.}(2020)\citenamefont {Chai} \emph
  {et~al.}}]{Chai:2020nxw}%
  \BibitemOpen
  \bibfield  {author} {\bibinfo {author} {\bibfnamefont {Y.}~\bibnamefont
  {Chai}} \emph {et~al.},\ }\href {\doibase 10.1103/PhysRevD.102.014508}
  {\bibfield  {journal} {\bibinfo  {journal} {Phys. Rev. D}\ }\textbf {\bibinfo
  {volume} {102}},\ \bibinfo {pages} {014508} (\bibinfo {year} {2020})},\
  \Eprint {http://arxiv.org/abs/2002.12044} {arXiv:2002.12044 [hep-lat]}
  \BibitemShut {NoStop}%
\bibitem [{\citenamefont {Dudek}\ \emph {et~al.}(2012)\citenamefont {Dudek}
  \emph {et~al.}}]{Dudek:2012vr}%
  \BibitemOpen
  \bibfield  {author} {\bibinfo {author} {\bibfnamefont {J.}~\bibnamefont
  {Dudek}} \emph {et~al.},\ }\href {\doibase 10.1140/epja/i2012-12187-1}
  {\bibfield  {journal} {\bibinfo  {journal} {Eur. Phys. J. A}\ }\textbf
  {\bibinfo {volume} {48}},\ \bibinfo {pages} {187} (\bibinfo {year} {2012})},\
  \Eprint {http://arxiv.org/abs/1208.1244} {arXiv:1208.1244 [hep-ex]}
  \BibitemShut {NoStop}%
\bibitem [{\citenamefont {Tanabashi}\ \emph {et~al.}(2018)\citenamefont
  {Tanabashi} \emph {et~al.}}]{Tanabashi:2018oca}%
  \BibitemOpen
  \bibfield  {author} {\bibinfo {author} {\bibfnamefont {M.}~\bibnamefont
  {Tanabashi}} \emph {et~al.} (\bibinfo {collaboration} {Particle Data
  Group}),\ }\href {\doibase 10.1103/PhysRevD.98.030001} {\bibfield  {journal}
  {\bibinfo  {journal} {Phys. Rev. D}\ }\textbf {\bibinfo {volume} {98}},\
  \bibinfo {pages} {030001} (\bibinfo {year} {2018})}\BibitemShut {NoStop}%
\bibitem [{\citenamefont {Bali}\ \emph {et~al.}(2016)\citenamefont {Bali},
  \citenamefont {Lang}, \citenamefont {Musch},\ and\ \citenamefont
  {Sch\"afer}}]{Bali:2016lva}%
  \BibitemOpen
  \bibfield  {author} {\bibinfo {author} {\bibfnamefont {G.~S.}\ \bibnamefont
  {Bali}}, \bibinfo {author} {\bibfnamefont {B.}~\bibnamefont {Lang}}, \bibinfo
  {author} {\bibfnamefont {B.~U.}\ \bibnamefont {Musch}}, \ and\ \bibinfo
  {author} {\bibfnamefont {A.}~\bibnamefont {Sch\"afer}},\ }\href {\doibase
  10.1103/PhysRevD.93.094515} {\bibfield  {journal} {\bibinfo  {journal} {Phys.
  Rev. D}\ }\textbf {\bibinfo {volume} {93}},\ \bibinfo {pages} {094515}
  (\bibinfo {year} {2016})},\ \Eprint {http://arxiv.org/abs/1602.05525}
  {arXiv:1602.05525 [hep-lat]} \BibitemShut {NoStop}%
\bibitem [{\citenamefont {Ji}\ \emph {et~al.}(2018)\citenamefont {Ji},
  \citenamefont {Zhang},\ and\ \citenamefont {Zhao}}]{Ji:2017oey}%
  \BibitemOpen
  \bibfield  {author} {\bibinfo {author} {\bibfnamefont {X.}~\bibnamefont
  {Ji}}, \bibinfo {author} {\bibfnamefont {J.-H.}\ \bibnamefont {Zhang}}, \
  and\ \bibinfo {author} {\bibfnamefont {Y.}~\bibnamefont {Zhao}},\ }\href
  {\doibase 10.1103/PhysRevLett.120.112001} {\bibfield  {journal} {\bibinfo
  {journal} {Phys. Rev. Lett.}\ }\textbf {\bibinfo {volume} {120}},\ \bibinfo
  {pages} {112001} (\bibinfo {year} {2018})},\ \Eprint
  {http://arxiv.org/abs/1706.08962} {arXiv:1706.08962 [hep-ph]} \BibitemShut
  {NoStop}%
\bibitem [{\citenamefont {Ishikawa}\ \emph {et~al.}(2017)\citenamefont
  {Ishikawa}, \citenamefont {Ma}, \citenamefont {Qiu},\ and\ \citenamefont
  {Yoshida}}]{Ishikawa:2017faj}%
  \BibitemOpen
  \bibfield  {author} {\bibinfo {author} {\bibfnamefont {T.}~\bibnamefont
  {Ishikawa}}, \bibinfo {author} {\bibfnamefont {Y.-Q.}\ \bibnamefont {Ma}},
  \bibinfo {author} {\bibfnamefont {J.-W.}\ \bibnamefont {Qiu}}, \ and\
  \bibinfo {author} {\bibfnamefont {S.}~\bibnamefont {Yoshida}},\ }\href
  {\doibase 10.1103/PhysRevD.96.094019} {\bibfield  {journal} {\bibinfo
  {journal} {Phys. Rev. D}\ }\textbf {\bibinfo {volume} {96}},\ \bibinfo
  {pages} {094019} (\bibinfo {year} {2017})},\ \Eprint
  {http://arxiv.org/abs/1707.03107} {arXiv:1707.03107 [hep-ph]} \BibitemShut
  {NoStop}%
\bibitem [{\citenamefont {Green}\ \emph {et~al.}(2018)\citenamefont {Green},
  \citenamefont {Jansen},\ and\ \citenamefont {Steffens}}]{Green:2017xeu}%
  \BibitemOpen
  \bibfield  {author} {\bibinfo {author} {\bibfnamefont {J.}~\bibnamefont
  {Green}}, \bibinfo {author} {\bibfnamefont {K.}~\bibnamefont {Jansen}}, \
  and\ \bibinfo {author} {\bibfnamefont {F.}~\bibnamefont {Steffens}},\ }\href
  {\doibase 10.1103/PhysRevLett.121.022004} {\bibfield  {journal} {\bibinfo
  {journal} {Phys. Rev. Lett.}\ }\textbf {\bibinfo {volume} {121}},\ \bibinfo
  {pages} {022004} (\bibinfo {year} {2018})},\ \Eprint
  {http://arxiv.org/abs/1707.07152} {arXiv:1707.07152 [hep-lat]} \BibitemShut
  {NoStop}%
\bibitem [{\citenamefont {Stewart}\ and\ \citenamefont
  {Zhao}(2018)}]{Stewart:2017tvs}%
  \BibitemOpen
  \bibfield  {author} {\bibinfo {author} {\bibfnamefont {I.~W.}\ \bibnamefont
  {Stewart}}\ and\ \bibinfo {author} {\bibfnamefont {Y.}~\bibnamefont {Zhao}},\
  }\href {\doibase 10.1103/PhysRevD.97.054512} {\bibfield  {journal} {\bibinfo
  {journal} {Phys. Rev. D}\ }\textbf {\bibinfo {volume} {97}},\ \bibinfo
  {pages} {054512} (\bibinfo {year} {2018})},\ \Eprint
  {http://arxiv.org/abs/1709.04933} {arXiv:1709.04933 [hep-ph]} \BibitemShut
  {NoStop}%
\bibitem [{\citenamefont {Chen}\ \emph {et~al.}(2018)\citenamefont {Chen},
  \citenamefont {Ishikawa}, \citenamefont {Jin}, \citenamefont {Lin},
  \citenamefont {Yang}, \citenamefont {Zhang},\ and\ \citenamefont
  {Zhao}}]{Chen:2017mzz}%
  \BibitemOpen
  \bibfield  {author} {\bibinfo {author} {\bibfnamefont {J.-W.}\ \bibnamefont
  {Chen}}, \bibinfo {author} {\bibfnamefont {T.}~\bibnamefont {Ishikawa}},
  \bibinfo {author} {\bibfnamefont {L.}~\bibnamefont {Jin}}, \bibinfo {author}
  {\bibfnamefont {H.-W.}\ \bibnamefont {Lin}}, \bibinfo {author} {\bibfnamefont
  {Y.-B.}\ \bibnamefont {Yang}}, \bibinfo {author} {\bibfnamefont {J.-H.}\
  \bibnamefont {Zhang}}, \ and\ \bibinfo {author} {\bibfnamefont
  {Y.}~\bibnamefont {Zhao}},\ }\href {\doibase 10.1103/PhysRevD.97.014505}
  {\bibfield  {journal} {\bibinfo  {journal} {Phys. Rev. D}\ }\textbf {\bibinfo
  {volume} {97}},\ \bibinfo {pages} {014505} (\bibinfo {year} {2018})},\
  \Eprint {http://arxiv.org/abs/1706.01295} {arXiv:1706.01295 [hep-lat]}
  \BibitemShut {NoStop}%
\bibitem [{\citenamefont {Alexandrou}\ \emph {et~al.}(2011)\citenamefont
  {Alexandrou}, \citenamefont {Constantinou}, \citenamefont {Korzec},
  \citenamefont {Panagopoulos},\ and\ \citenamefont
  {Stylianou}}]{Alexandrou:2010me}%
  \BibitemOpen
  \bibfield  {author} {\bibinfo {author} {\bibfnamefont {C.}~\bibnamefont
  {Alexandrou}}, \bibinfo {author} {\bibfnamefont {M.}~\bibnamefont
  {Constantinou}}, \bibinfo {author} {\bibfnamefont {T.}~\bibnamefont
  {Korzec}}, \bibinfo {author} {\bibfnamefont {H.}~\bibnamefont
  {Panagopoulos}}, \ and\ \bibinfo {author} {\bibfnamefont {F.}~\bibnamefont
  {Stylianou}},\ }\href {\doibase 10.1103/PhysRevD.83.014503} {\bibfield
  {journal} {\bibinfo  {journal} {Phys. Rev. D}\ }\textbf {\bibinfo {volume}
  {83}},\ \bibinfo {pages} {014503} (\bibinfo {year} {2011})},\ \Eprint
  {http://arxiv.org/abs/1006.1920} {arXiv:1006.1920 [hep-lat]} \BibitemShut
  {NoStop}%
\bibitem [{\citenamefont {Izubuchi}\ \emph {et~al.}(2018)\citenamefont
  {Izubuchi}, \citenamefont {Ji}, \citenamefont {Jin}, \citenamefont
  {Stewart},\ and\ \citenamefont {Zhao}}]{Izubuchi:2018srq}%
  \BibitemOpen
  \bibfield  {author} {\bibinfo {author} {\bibfnamefont {T.}~\bibnamefont
  {Izubuchi}}, \bibinfo {author} {\bibfnamefont {X.}~\bibnamefont {Ji}},
  \bibinfo {author} {\bibfnamefont {L.}~\bibnamefont {Jin}}, \bibinfo {author}
  {\bibfnamefont {I.~W.}\ \bibnamefont {Stewart}}, \ and\ \bibinfo {author}
  {\bibfnamefont {Y.}~\bibnamefont {Zhao}},\ }\href {\doibase
  10.1103/PhysRevD.98.056004} {\bibfield  {journal} {\bibinfo  {journal} {Phys.
  Rev. D}\ }\textbf {\bibinfo {volume} {98}},\ \bibinfo {pages} {056004}
  (\bibinfo {year} {2018})},\ \Eprint {http://arxiv.org/abs/1801.03917}
  {arXiv:1801.03917 [hep-ph]} \BibitemShut {NoStop}%
\bibitem [{\citenamefont {Fan}\ \emph {et~al.}(2020)\citenamefont {Fan},
  \citenamefont {Gao}, \citenamefont {Li}, \citenamefont {Lin}, \citenamefont
  {Karthik}, \citenamefont {Mukherjee}, \citenamefont {Petreczky},
  \citenamefont {Syritsyn}, \citenamefont {Yang},\ and\ \citenamefont
  {Zhang}}]{Fan:2020nzz}%
  \BibitemOpen
  \bibfield  {author} {\bibinfo {author} {\bibfnamefont {Z.}~\bibnamefont
  {Fan}}, \bibinfo {author} {\bibfnamefont {X.}~\bibnamefont {Gao}}, \bibinfo
  {author} {\bibfnamefont {R.}~\bibnamefont {Li}}, \bibinfo {author}
  {\bibfnamefont {H.-W.}\ \bibnamefont {Lin}}, \bibinfo {author} {\bibfnamefont
  {N.}~\bibnamefont {Karthik}}, \bibinfo {author} {\bibfnamefont
  {S.}~\bibnamefont {Mukherjee}}, \bibinfo {author} {\bibfnamefont
  {P.}~\bibnamefont {Petreczky}}, \bibinfo {author} {\bibfnamefont
  {S.}~\bibnamefont {Syritsyn}}, \bibinfo {author} {\bibfnamefont {Y.-B.}\
  \bibnamefont {Yang}}, \ and\ \bibinfo {author} {\bibfnamefont
  {R.}~\bibnamefont {Zhang}},\ }\href {\doibase 10.1103/PhysRevD.102.074504}
  {\bibfield  {journal} {\bibinfo  {journal} {Phys. Rev. D}\ }\textbf {\bibinfo
  {volume} {102}},\ \bibinfo {pages} {074504} (\bibinfo {year} {2020})},\
  \Eprint {http://arxiv.org/abs/2005.12015} {arXiv:2005.12015 [hep-lat]}
  \BibitemShut {NoStop}%
\bibitem [{\citenamefont {Constantinou}\ and\ \citenamefont
  {Panagopoulos}(2017)}]{Constantinou:2017sej}%
  \BibitemOpen
  \bibfield  {author} {\bibinfo {author} {\bibfnamefont {M.}~\bibnamefont
  {Constantinou}}\ and\ \bibinfo {author} {\bibfnamefont {H.}~\bibnamefont
  {Panagopoulos}},\ }\href {\doibase 10.1103/PhysRevD.96.054506} {\bibfield
  {journal} {\bibinfo  {journal} {Phys. Rev. D}\ }\textbf {\bibinfo {volume}
  {96}},\ \bibinfo {pages} {054506} (\bibinfo {year} {2017})},\ \Eprint
  {http://arxiv.org/abs/1705.11193} {arXiv:1705.11193 [hep-lat]} \BibitemShut
  {NoStop}%
\bibitem [{\citenamefont {Radyushkin}(2018)}]{Radyushkin:2017lvu}%
  \BibitemOpen
  \bibfield  {author} {\bibinfo {author} {\bibfnamefont {A.}~\bibnamefont
  {Radyushkin}},\ }\href {\doibase 10.1016/j.physletb.2018.04.023} {\bibfield
  {journal} {\bibinfo  {journal} {Phys. Lett. B}\ }\textbf {\bibinfo {volume}
  {781}},\ \bibinfo {pages} {433} (\bibinfo {year} {2018})},\ \Eprint
  {http://arxiv.org/abs/1710.08813} {arXiv:1710.08813 [hep-ph]} \BibitemShut
  {NoStop}%
\bibitem [{\citenamefont {Chen}\ \emph
  {et~al.}(2016{\natexlab{b}})\citenamefont {Chen}, \citenamefont {Cohen},
  \citenamefont {Ji}, \citenamefont {Lin},\ and\ \citenamefont
  {Zhang}}]{Chen:2016utp}%
  \BibitemOpen
  \bibfield  {author} {\bibinfo {author} {\bibfnamefont {J.-W.}\ \bibnamefont
  {Chen}}, \bibinfo {author} {\bibfnamefont {S.~D.}\ \bibnamefont {Cohen}},
  \bibinfo {author} {\bibfnamefont {X.}~\bibnamefont {Ji}}, \bibinfo {author}
  {\bibfnamefont {H.-W.}\ \bibnamefont {Lin}}, \ and\ \bibinfo {author}
  {\bibfnamefont {J.-H.}\ \bibnamefont {Zhang}},\ }\href {\doibase
  10.1016/j.nuclphysb.2016.07.033} {\bibfield  {journal} {\bibinfo  {journal}
  {Nucl. Phys. B}\ }\textbf {\bibinfo {volume} {911}},\ \bibinfo {pages} {246}
  (\bibinfo {year} {2016}{\natexlab{b}})},\ \Eprint
  {http://arxiv.org/abs/1603.06664} {arXiv:1603.06664 [hep-ph]} \BibitemShut
  {NoStop}%
\bibitem [{\citenamefont
  {Radyushkin}(2017{\natexlab{b}})}]{Radyushkin:2017ffo}%
  \BibitemOpen
  \bibfield  {author} {\bibinfo {author} {\bibfnamefont {A.}~\bibnamefont
  {Radyushkin}},\ }\href {\doibase 10.1016/j.physletb.2017.05.024} {\bibfield
  {journal} {\bibinfo  {journal} {Phys. Lett. B}\ }\textbf {\bibinfo {volume}
  {770}},\ \bibinfo {pages} {514} (\bibinfo {year} {2017}{\natexlab{b}})},\
  \Eprint {http://arxiv.org/abs/1702.01726} {arXiv:1702.01726 [hep-ph]}
  \BibitemShut {NoStop}%
\bibitem [{\citenamefont {Braun}\ \emph {et~al.}(1995)\citenamefont {Braun},
  \citenamefont {Gornicki},\ and\ \citenamefont {Mankiewicz}}]{Braun:1994jq}%
  \BibitemOpen
  \bibfield  {author} {\bibinfo {author} {\bibfnamefont {V.}~\bibnamefont
  {Braun}}, \bibinfo {author} {\bibfnamefont {P.}~\bibnamefont {Gornicki}}, \
  and\ \bibinfo {author} {\bibfnamefont {L.}~\bibnamefont {Mankiewicz}},\
  }\href {\doibase 10.1103/PhysRevD.51.6036} {\bibfield  {journal} {\bibinfo
  {journal} {Phys. Rev. D}\ }\textbf {\bibinfo {volume} {51}},\ \bibinfo
  {pages} {6036} (\bibinfo {year} {1995})},\ \Eprint
  {http://arxiv.org/abs/hep-ph/9410318} {arXiv:hep-ph/9410318} \BibitemShut
  {NoStop}%
\end{thebibliography}%

\end{document}